\documentclass{aa}
\usepackage{graphicx}
\usepackage{txfonts}

\begin{document}
\title{Multi wavelength study of the gravitational lens system RXS~J1131-1231}

\subtitle{II Lens model and source reconstruction\thanks{Based on observations made with the NASA/ESA Hubble Space Telescope, obtained at the Space Telescope Science Institute, which is operated by the Association of Universities for Research in Astronomy, Inc., under NASA contract NAS 5-26555. Present observations were made under HST-GO-9744 (P.I. C.S. Kochanek).}}

\author{
  J.-F. Claeskens\inst{1},
  D. Sluse\inst{1,2},
  P. Riaud\inst{1},
  \and
  J. Surdej\inst{1}\thanks{Directeur de Recherches honoraire du F.N.R.S. (Belgium)}.
  }

\institute{Institut d'Astrophysique et de G{\'e}ophysique,
  Universit{\'e} de Li{\`e}ge, All{\'e}e du 6 Ao{\^u}t 17, 
  B-4000 Sart Tilman (Li{\`e}ge), Belgium \and 
  Laboratoire d'Astrophysique, Ecole Polytechnique F\'ed\'erale de Lausanne  (EPFL) Observatoire, CH-1290 Sauverny, Switzerland}
  \offprints{J.-F. Claeskens
  \email{claesken@astro.ulg.ac.be}
}
\date{}

\abstract{}
{High angular resolution images of the complex gravitational lens system \object{RXS~J1131-1231} (a quadruply imaged AGN with a bright Einstein ring) obtained with the Advanced Camera for Surveys and NICMOS instruments onboard the Hubble Space Telescope are analysed to determine the lens model and to reconstruct the host galaxy.} 
{ The lens model is constrained by the relative astrometric positions of the lens and point-like images, and by the extended lensed structures. The \textit{non-parametric} light distribution is recovered in the source plane by means of back ray-tracing.}
{1- Precise astrometry and photometry of the four QSO lensed images (A-D) and of the lensing galaxy (G) are obtained.  They are found  in agreement with an independent study presented in a companion paper. The position and colours of the X object seen in projection close to the lens are found to be only compatible with a satellite galaxy associated with the lens. 

2- The  Singular Isothermal Ellipsoid  plus external shear provides a good fit of the astrometry of images A-D. The positions of extended substructures are also well reproduced. However an octupole ($m=4$) must be added to the lens potential in order to reproduce the observed lens position, as well as the $I_{\rm B}/I_{\rm C}$ point-like image flux ratio. The ellipticity and orientation of the mass quadrupole are found similar to those of the light distribution, fitted by a S\'ersic profile. The lens ($z=0.295$) is found to be a massive elliptical in a rich environment and showing possible evolution with respect to $z=0$. 

3-  The host galaxy ($z=0.658$) is found to be a substantially magnified ($M\sim 9$) luminous Seyfert 1 spiral galaxy. The angular resolution is sufficient to see regions where stars are intensively forming. Interaction with a closeby  companion is also observed. 

4- Finally, in the case of \object{RXS~J1131-1231}, extended lensed structures do not  help much in constraining the lens model. }
{}


\titlerunning{RXS J1131-1231: Lens model and source reconstruction}
\authorrunning{Claeskens et al.}
\maketitle
%

\section{Introduction}
\label{sec:intro}

\object{RXS~J1131-1231} (hereafter J1131) was discovered by Sluse et al. (2003\cite{sluse03}) as a quasar located at  $z_s =0.658$ and quadruply imaged by a bright elliptical galaxy located at $z_l=0.295$. A thorough study of the observed flux ratios at different epochs and in several bands has been presented in Sluse et al. (2006\cite{sluse06}; paper I). 

The QSO host galaxy is also lensed and appears as a nearly complete ($\sim 305^\circ$) bright Einstein ring, clearly seen not only in the near infra-red, but also at visible wavelengths (angular radius $\theta_E \sim 1.8''$). Thanks to the ``proximity'' of the source, the ring is much brighter than in other known lens systems showing  extended structures, such as \object{MG0414+0534} (Falco et al. 1997\cite{falco97}), \object{PG1115+080} (Impey et al. 1998\cite{impey98}),  \object{ER0047-2808} (Wayth et al. 2005\cite{wayth05}, Dye and Warren 2005\cite{dye05}) or  \object{FOR J0332-3557} (Cabanac et al. 2005\cite{cabanac05}). Impressive images obtained with the Advanced Camera for Survey (ACS) and NICMOS onboard the Hubble Space Telescope (HST) are presented and analysed in this paper. Besides the Einstein ring,  they reveal detailed arcs, arclets and substructures in the ring (see Fig.\ref{fig:rgb}), corresponding to patchy emission in the host galaxy. Since Kochanek et al. (2001\cite{koch01}) state that {\it ``the shape of an Einstein ring accurately and independently determines the shape of the lens potential and the shape of the lensed host galaxy''}, such images should thus represent a potential golden mine to constrain the lens model at best as well as to recover the morphology of a QSO host at $z=0.66$ with an unprecedented angular resolution. Let us first investigate in this introduction what we can really expect.

\subsection{Constraining the lens model}

It is well known that constraints provided only by the relative astrometry of the lensed images\footnote{Observed flux ratios must be corrected for differential  extinction, delayed intrinsic variability and/or microlensing before being used to constrain a lens model.} of a single point-like source are not sufficiently numerous to avoid degeneracies in the lens potential, which can then lead to wrong estimates of the $H_o$ value derived from time delay measurements (for a general discussion on degeneracies, see e.g. Saha 2000\cite{saha00}, Gorenstein et al. 1988\cite{goren88}). Concentrating on the mass distribution, three kinds of degeneracy can be identified: the mass-sheet degeneracy (e.g. Saha 2000\cite{saha00}), the radial mass-profile degeneracy (e.g. Refsdal \& Surdej 1994\cite{refsdal94} in the circular case and Wucknitz 2002\cite{wucknitz02} in a more general case) and the degeneracy between the internal ellipticity, the external shear strength and their relative orientation (Witt \& Mao 1997\cite{witt97}, Keeton et al. 1997\cite{keeton97}).

How can extended lensed structures help in avoiding those degeneracies?  First of all, since the mass-sheet degeneracy implies a simple rescaling of the lens mass without affecting the image structure, the presence of an Einstein ring will {\it not} break it, as pointed out by Treu and Koopmans (2002\cite{treukoop02}) in their analysis of PG1115+080. Indeed, this degeneracy would be broken with lensing data only if sources at different redshifts were simultaneously lensed (because the critical surface mass density, and thus the scaling,  changes with redshift). 

Second, the mass-profile degeneracy. Knowing the exact value of the slope $\beta$ of the mass distribution ($\kappa \propto r^{-\beta}$) is required to derive the correct value of $H_o$. Arbitrarily fixing $\beta=1$ (i.e. isothermal distribution) may lead to a systematic error of 10\% (Wucknitz 2002\cite{wucknitz02}). Despite the fact that the observed $\beta$ value may be affected by the mass-sheet degeneracy (Wucknitz \& Refsdal 2001\cite{wucknitz01}), it should thus be determined from the lensing data. Extended structures may help at this point because astrometric constraints alone are unable to do so, even when the number of degrees of freedom is sufficient. Indeed any astrometry will be identically reproduced with no external shear and a sheet of constant, critical density ($\beta=0$). A second point-like source at least, or extended structures is/are necessary to raise this pathology. Now, in view of testing theoretical mass radial profiles (such as the halo profile predicted in the Cold Dark Matter by Navarro, Frenk and White 1996\cite{navarro96}, 1997\cite{navarro97}) one should bear in mind that gravitational lensing (GL) can only constrain the radial profile within a ring defined by the distances of the most external and the most internal images (Kochanek 2005\cite{koch05}). In this context, the cusp configuration of J1131 is  more favourable than the more symmetrical cross configuration (like \object{H1413+117} for example) but in such a case the extended Einstein ring lies in the radial interval already probed  by the point-like images (see Fig. \ref{fig:rgb}) and thus, does not help much. The ideal probe would in fact be a second point-like source, located just behind the lens, and giving rise to a smaller Einstein ring radius because of geometric considerations.  However, stellar dynamics also appears as a useful complement to GL to put stronger constraints on the mass profile (Treu \& Koopmans 2004\cite{treu04}). 

Third, the external shear, the internal ellipticity of the mass distribution and their relative orientation can combine in several ways to produce a given net potential quadrupole. An asymmetric lensing configuration like J1131 (see also \object{B1422+231} in Keeton et al. 1997\cite{keeton97}) and additional azimuthal constraints provided by extended structures may help in reducing this type of degeneracy.   

Finally, two remarks concerning the constraints provided by extended structures. First, they are less robust than the astrometric constraints since they rely on the assumed surface brightness conservation law. Second, not all the image pixels carry independent information\footnote{Saha and Williams (2001\cite{saha01}) have shown that the global shape of the Einstein ring is encoded into the point-like image time delay ratios.}. Indeed, obviously, extended structures are relevant to constrain the mass model only when different images can be mapped onto each other. Thus, {\it a priori}, the cusp configuration of J1131 is {\it not} favourable in this context since the three brightest images are merging and cannot be compared over a spatial extent much larger than the PSF (in contrast with the case of \object{Q0957+561}, Keeton et al. 2000\cite{keeton00}). In the language of isophotal separatrices (Suyu \& Blandford 2006\cite{suyu06}; see also Kochanek et al. 2001\cite{koch01}), the strong magnification of a very small region in the source plane produces a nearly constant surface brightness arc where it is impossible to precisely identify isophotes or flux minima.

Hopefully, in the case of J1131, lensed extended {\it sub}-structures are clearly identified on the HST images and are {\it a priori} useful as further constraints on the lens model.

\subsection{The lens fundamental plane}
Another important feature of J1131 is the fact that the lensing galaxy is bright and rather well separated from the lensed structures on the HST images. This makes possible a good determination of its photometric properties. Furthermore, Sluse et al. (2003\cite{sluse03}) have already shown that the lens is an elliptical galaxy located at $z_l =0.295$. Grapping informations on both the mass and the light properties of a non-local elliptical galaxy allows to probe the evolution of the M/L ratio and the stellar formation history through the position of the galaxy with respect to the so-called local {\it Fundamental Plane} (FP). The latter is a relation between the velocity dispersion, the effective radius and the surface brightness of elliptical galaxies (Dressler et al. 1987\cite{dressler87}; Djorgovski and Davis 1987\cite{djorgov87}; J\o rgensen, Franx and Kj\ae rgaard 1995a\cite{jorg95a}, 1995b\cite{jorg95b}, 1996\cite{jorg96}). 

Isolated field ellipticals, selected on the basis of their morphology and colours from the Medium Deep Survey  up to $z\sim0.7$, have been compared to the local FP (Treu et al. 2001\cite{treu01}; Treu et al. 2002\cite{treu02}) and an evolution of the M/L ratio has been found in the sense that distant galaxies are more luminous for a given effective radius and velocity dispersion. 

 More recently, deep spectroscopy with Keck and VLT has allowed to probe the evolution of field early type galaxies up to $z\sim 1$ (di Serego Alighieri et al. 2005\cite{alighieri05}, van der Wel et al. 2005\cite{vanderwel05}, Treu et al. 2005a\cite{treu05a},b\cite{treu05b}). A common trend is that the evolution is mass-dependent: more massive galaxies evolve more slowly than less massive ones, probably because the latter have significantly younger stellar populations. In fact the evolution rate of massive field galaxies is similar to that of cluster galaxies (van der Wel et al. 2005\cite{vanderwel05}) and mass rather than environment is found to drive the evolution (Treu et al. 2005a\cite{treu05a}).

Gravitational lenses offer a completely different selection criterium, i.e. based on mass instead of luminosity, and also to redshifts up to $z\simeq 1$. A first study by Kochanek et al. (2000\cite{koch00}) has been followed by the works by Rusin et al. (2003\cite{rusin03}) and van de Ven et al. (2003\cite{vandeven03}), which confirm a statistical evolution of the M/L ratio in the B and r bands. Since the data presented in this paper allow to estimate the velocity dispersion from the lens model fitting and the morphological parameters from the fit of the observed surface brightness profile, it will be possible to check whether the lens of J1131 belongs to the local FP or whether significant evolution is present.

\subsection{The host galaxy}

Since only the {\it matching} between the lensed images is used to constrain the lens model but not the {\it real shape} of the source, the latter can also be {\it independently} reconstructed. The study of the host galaxy and its environment at $z=0.658$ is thus possible.

Such studies are interesting because host galaxies and their environment are closely related to the AGN phenomenon itself. The present view of galactic activity requires two components: a central engine -- a supermassive black hole -- and fueling gas feeding the engine through accretion processes (e.g. Antonucci 1993\cite{anton93}). Even if {\it ``black holes have been discovered in every {\rm [local]} galaxy that contains a bulge\footnote{And the black hole mass correlates with the {\it bulge} luminosity, not with the total host luminosity (e.g. Magorrian et al. 1998\cite{magorrian98}).} and that has been observed with enough resolution''} (Kormendy and Gebhardt 2001\cite{kormendy01}), each galactic bulge does not harbour an AGN. Therefore, the limiting condition for nuclear activity seems to be gas fueling.

From the morphological point of view, the classical picture is that the majority of AGN are hosted by early-type, bulge dominated galaxies (i.e. 65 - 85\% according to S\'anchez et al. 2004, Dunlop et al. 2003\cite{dunlop03}) and that disk dominated gas-rich galaxies only host small black holes (Kormendy \& Gebhardt 2001\cite{kormendy01}), and  weak $M_V > -23$ AGN (Dunlop et al. 2003\cite{dunlop03}). The feeding (through the AGN phase) and growth of the central black hole thus seems related to the bulge history rather than to the disk gas content.  A coherent explanation relies then on  merging and galaxy interactions (e.g. Di Matteo et al. 2005\cite{dimatteo05}). Besides triggering stellar formation, merging would supply fresh gas to the black hole whose activity would be regulated by its own energy release and the gravitational potential of the spheroid. If such is the case, AGN should statistically correlate with the presence of stellar formation in the host and galaxy interactions.

S\'anchez et al. (2004) indeed find that roughly 70\% of the morphologically early-type hosts have blue rest frame colours. This is significantly larger than in inactive galaxies at the same redshift ($0.5 < z < 1.1$). Focusing on the luminosity of the [O {\sc iii}] line, Kauffmann et al. (2003\cite{kauff03}) find that highly luminous AGN with $0.02<z<0.3$ are hosted in bulges with young stellar populations. However, Grogin et al. (2005\cite{grogin05}) and S\'anchez et al. (2004) do not observe a significantly higher fraction of merging among active galaxies than among inactive ones. The latter authors argue that the morphological signature of interaction may have already disappeared while the bluer colour is still being seen. Finally,  Dunlop et al. (2003)\cite{dunlop03} claim that the apparent lack of enhanced merging they observe in active galaxies is not a problem since an accretion rate of only 1 M$_\odot$ yr$^{-1}$ is enough to activate a quasar. Obviously, the debate is still open.

From an observational point of view, determining the luminosity profile and colours of the host as well as the signs of merging requires high angular resolution. The bright PSF associated with the unresolved AGN makes the task difficult. Even with the narrow HST PSF, only relatively low redshift hosts have been investigated so far (typically $z< 0.5$). The $(1+z)^4$ cosmological surface brightness dimming joined to the bluer rest frame spectra observed in red filters make the contrast between QSOs and hosts even worst with increasing $z$. Thanks to NICMOS observations of $z\simeq 1$ QSOs, Kukula et al. (2001\cite{kukula01}) claim that hosts statistically follow the same Kormendy relation as 3CR galaxies at the same redshift, while at $z\simeq 2$, only the total host luminosity of the host could be determined within a factor 2.

By stretching the host images, gravitational lensing helps in revealing host structures located closer to the QSO and does improve the resolution in the source plane (although the improvement is limited and not isotropic, see Sect. \ref{sec:source}). The opportunity to recover the host profile at higher redshifts is thus offered for the case of J1131. A larger S/N due to the larger number of collected photons is also expected. However, contrary to some believes, lensing does not decrease the contrast between the central QSO and the host (the host surface brightness is preserved while the flux of the point-like images are {\it amplified} since the latter are not resolved!), so that the noise associated with the  central PSF residuals is not smaller than for unlensed objects. 

A first statistical study of lensed host galaxies with $z > 1.5$ found in the CASTLES gravitational lensing survey has been done by Peng et al. (2004\cite{peng04}).  Most of them are luminous and  some of them also show blue colours possibly betraying star formation in the QSO rest frame.

In the case of J1131, located at $z=0.658$, the increased angular resolution offered by the combination of HST+ACS and gravitational lensing helps in recovering photometric substructures (e.g. star forming regions) in the host as well as possible signatures of merging and/or interaction. In order to keep those potentialities, the source must be retrieved in a model independent way. This is done with the technique of back ray-tracing. This implies no fitting of any source model or source grid intensity, while saving much computing time.

\vspace{0.5cm}

 In the following, we first describe the observations and their reduction in Sec. \ref{sec:obs}, then we present our technique to reconstruct the source in Sect. \ref{sec:source}. In Sect. \ref{sec:fitting}, we detail the successive steps followed to fit the lens system parameters and we give the main results. The latter are discussed in terms of the lens and host properties in Sect. \ref{sec:result}.  Finally a summary on the morphology of this complex gravitational lens system is given in Sect. \ref{sec:conclusion}.

Throughout this paper, we use the cosmological parameter values $\Omega_o=0.3$, $\lambda_o=0.7$ and $H_o = 50 h_{\rm 50}$ km/s/Mpc.

\section{Observations, reduction and photometric calibrations}
\label{sec:obs}

\begin{table*}[t]
\begin{center}
\caption{Log of observations for RXS J1131-1231 with {\it HST}.}
\label{tab:obs}
\begin{tabular}{cccccccc}
\hline
Date & Instrument & Filter & NExp & Exp(s) & Scale ($''$/pix) & RON (e/pix)& g (e/ADU)\\
\hline
17-11-2003 & NICMOS-NIC2 & F160W & 5 & 640 & 0.0759/0.0754$^\dag$ & 26 & 5.4 \\
17-11-2003 & NICMOS-NIC2 & F160W & 3 & 704 & 0.0759/0.0754$^\dag$ & 26 & 5.4 \\
22-06-2004 & ACS-WFC & F555W & 5 & 396 & 0.050$^\ddag$ &5.3 & 2.0\\
24-06-2004 & ACS-WFC & F814W & 5 & 396 & 0.050$^\ddag$ &5.3 & 2.0\\
\hline
\multicolumn{8}{l}{$^\dag$ NIC2 plate scale along X and Y respectively (see http://www.stsci.edu/hst/nicmos/performance/platescale)}\\
\multicolumn{8}{l}{$^\ddag$ the reported ACS scale is obtained after drizzling}
\end{tabular}
\end{center}
\end{table*}

\subsection{Observations}

All observations have been obtained with the {\it Hubble Space Telescope} ({\it HST}) in the context of the CfA-Arizona Space Telescope Lens Survey (CASTLES). NIR imaging with NIC-2 camera has been obtained in filter F160W on 17-Nov-2003 and ACS images in F814W and F555W have been collected with the Wide Field Camera (WFC) on 22 and 24-June-2004, respectively. For each filter, several dithered frames have been obtained. The observation log is summarized in Table \ref{tab:obs}. 

\subsection{Reduction}
\label{sec:red}

The reduction procedure of the NIC-2 data has been described Paper I. Two frames have been discarded due to error flags in the brightest point-like images, leading to an effective value of NExp = 6 in F160W. On the other hand, a systematic error on the relative astrometry of 0.003$''$ has been adopted (Impey et al. 1998\cite{impey98}).

\vspace{0.25cm}
Regarding the ACS images, they have been individually reduced in 3 steps:

\begin{enumerate}
\item
Standard reduction (bias and dark subtraction, flatfield and gain corrections) has been performed with the STScI {\sc CALACS} pipeline.
\item
Cosmic ray cleaning has been performed on each individual frames with an  IDL procedure.  For each frame a mask of bad pixels has been built from cross-correlation with the other aligned frames. The affected pixel values have been replaced by the median of the four remaining frames. Neighbour pixels suffering from charge scattering (i.e. with value above the 5 $\sigma$ noise level) have also had their intensity replaced by a bilinear interpolation. We checked that no pixel has been replaced in the core of the PSFs. 
\item
The geometric image distortion of the ACS WFC had finally to be corrected because it strongly affects the photometry and astrometry of the sources as a function of their positions on the CCD. To reach that goal, we applied the {\sc STSDAS} (V3.3) task {\it Pydrizzle} (V3.3), which makes use of a fourth order polynomial model of the distortion.  This model is adequate for characterizing the distortion to an accuracy better than 0.1-0.2 pixels over the entire field of view (Pavlovsky et al. 2005\cite{pavlovsky05}). However, since J1131 does only extend over a few arcsec, we may expect a lower systematic error on the relative astrometry due to the geometric image distortion. Indeed, we checked on the DGEO distortion residual images corresponding to filters F555W and F814W provided by the STScI that, in the restricted detector areas where the lens is observed, the dispersion of the residual ($DX,DY$)  does not exceed 0.004 pixel around the mean value. Given the ACS scale, adopting a systematic error of 0.001$''$ on the {\it relative} astrometry is thus conservative since it is 3.5 larger than the expected error due to distortion residuals.

\begin{figure}
\begin{center}
\includegraphics[width=\columnwidth]{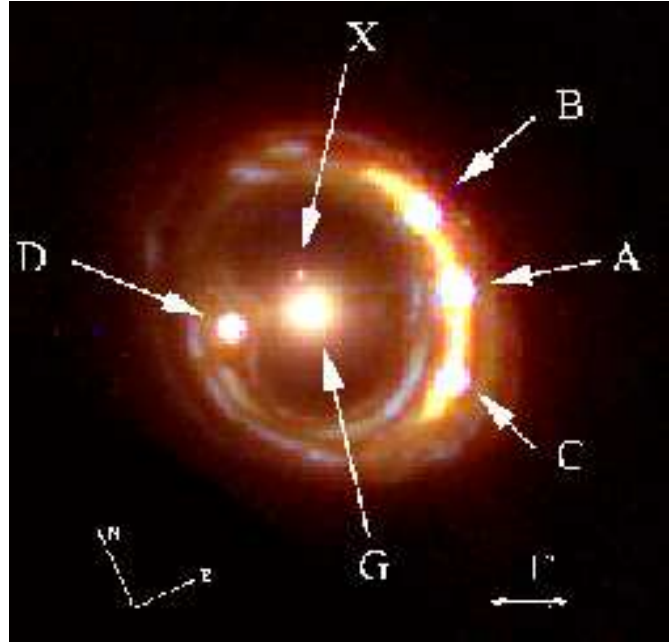}
\caption{Colour image of RXS J1131-1231 resulting from the combination of ACS F555W (blue), ACS F814W (green) and NICMOS2 F160W (red) images.}
\label{fig:rgb}
\end{center}
\end{figure}

One should note that interpolation might also produce a slight astrometric error but we think that it is included in the frame-to-frame measurement dispersion.
\end{enumerate}

For each frame, a synthetic PSF has been generated  with the Tiny Tim v 6.3 software (Krist \& Hook 2004\cite{krist04}) at the position of the lens centroid. For ACS PSFs, we noticed that Tiny Tim was not properly performing the geometric distortion correction, so that we processed the distorted PSFs with {\it Pydrizzle}, the same way as the images. It is worth noting that these final ACS PSFs had to be normalized manually.

A colour picture of  RXS J1131-1231 made by combining the reduced and coadded images in filters F555W, F814W and F160W is displayed in Fig. \ref{fig:rgb}. Besides the big yellow arc which is formed by 3 merging images of the most central part of the host, several thin arcs and arclets appear in blue and green colours: they are lensed images of extended substructures in the host galaxy (see Sect. \ref{sec:fit6}).

\subsection{Photometric calibrations}
\label{sec:photcal}

Absolute flux measurements are reported in the Vega magnitude system according to the following relations:

\begin{equation}
m_{\rm (F160W)} = -2.5 \log[{\rm PHOTNU}*CR*F^{-1}_\nu({\rm Vega})]\,,
\end{equation}
\begin{equation}
m_{\rm (F555W,F814W)} = -2.5 \log[CR] + m_{o}
\end{equation}

\noindent
where $CR$ is the count rate, {\sc photnu} = 1.49816e-6 Jy.Sec.DN$^{-1}$, $F_\nu({\rm Vega}) = 1043.5$ Jy, $m_o = 25.724$ for F555W and $m_o = 25.501$ for F814W, as provided on the NICMOS and ACS updated Web pages. A systematic calibration error is estimated to be 0.03 mag for NICMOS from the same Web page, while we adopt a formal value of 0.01 mag for ACS. 

Quality flags reveal that the peaks of QSO images A \& B are saturated on all F555W and F814W images. However the saturation is very light because visual inspection reveals that the PSF profile is not altered and masking the central pixels does not modify the PSF fitting. We are thus confident in our flux measurements reported below and made by means of PSF fittings.

\section{Image reconstruction in the source plane }
\label{sec:source}

The first source reconstruction algorithms have been developed in the radio domain when lensed extended radio structures were the only ones to be resolved. {\sc LensClean} was built on the basis of the Clean algorithm by Kochanek and Narayan (1992\cite{koch92}) and has recently been revisited by Wucknitz (2004\cite{wucknitz04}). On the other hand, the {\sc LensMem} algorithm was based on Maximum Entropy reconstruction (Wallington et al. 1996\cite{wallington96}). In these techniques, the shape of the source and the lens model are determined in two nested fitting procedures. More recently, in their so-called semi-linear inversion method, Warren and Dye (2004) replaced the source fitting by a matrix inversion. The two former algorithms proceed in the direction of propagation of light rays while the latter goes in the reverse direction (back ray-tracing) and, indeed, produces a deconvolved image in the source plane. Brewer and Lewis (2005\cite{brewer05}) have followed a more original approach based on genetic algorithms.

Let us assume now that the lens model is frozen. As in the algorithm proposed by Schramm \& Kayser (1987\cite{schramm87}), our method is based on  back ray-tracing and does not rely on any parametrized shape of the source. The technique of back ray-tracing is easy since it simply makes use of the direct lens equation:

\begin{equation}
\left\{ \begin{array}{rll}
x_s &=& x_i - \hat{\alpha_x}(x_i,y_i)\\
y_s &=& y_i - \hat{\alpha_y}(x_i,y_i)
	\end{array}
\right.\,,  \label{eq:lens} 
\end{equation}

\noindent
to find the (unique) source pixel $(x_s,y_s)$ corresponding to the image pixel $(x_i,y_i)$ given the displacement angle $(\hat{\alpha_x},\hat{\alpha_y})$. The source pixel intensity $I_s$ is defined as the observed intensity $I_i$, since the surface brightness is assumed to be preserved. Note that if several image pixels are mapped onto the same source pixel, the median value of the corresponding pixel values is to be taken  to remove the effect of deviant pixels (e.g. remaining after the PSF subtractions). The PSFs located where the unresolved QSO images are formed must indeed be removed since they do not satisfy the surface brightness conservation law. This source reconstruction algorithm is thus direct and does not need any fitting, any regularisation and is {\it non parametric}. The drawback is that the convolution process is not included. This is not a problem with the present high angular resolution HST images, but if the seeing were of the order of the ring width (about $1.5''$ for J1131) a preliminary deconvolution of the images would then be necessary. One should note however that deconvolution usually produces correlation between the intensities of neighbour pixels and artifacts in the background. The latter would partially compromise a clean source reconstruction.

In practice, our source reconstruction algorithm is based on the triangular pixel mapping described in Schneider, Ehlers \& Falco (1992\cite{schneider92}). Each squared image pixel is divided into two triangles which are then mapped onto the source plane with the lens equation (\ref{eq:lens}) and the deflection angle associated with the appropriate lens model. For each source pixel, a vector is filled with the intensities of all the image pixels mapped onto that pixel. We added an original option. It consists in selecting image pixels on the the basis of the local magnification value $M(i,j)$ in the image plane. Indeed, selecting only  image pixels with $M(i,j) \geq 1$ allows to ignore the flux observed inside the critical line, and thus, to  reduce the contaminating flux from the lens and to ignore the demagnified regions whose mapping leads to large noisy pixels in the source plane. This is a key point to make the sequential analysis of the lens system presented in Sect. \ref{sec:fitting}. Note that $M < 1$ pixels do not carry critical information since their values either are affected by the PSF residuals (image A) or correspond to averages over larger source regions and do not contain high resolution spatial information on the source. $M < 1$ pixels are not needed to reconstruct the image of the source since each of them has at least one corresponding $M>1$ pixel.

The choice of the pixel size in the source plane is not critical in the sense that the intensities in the source pixels are not considered as parameters (like in other fitting methods), but rather as quantities derived from the data (see also Sect. \ref{sec:fitting}). It is of course natural to choose a size smaller in the source plane than in the image plane since the whole source has been magnified. However, it is useless to consider an extremely small pixel size by adopting a $1/\sqrt{\mu}$ scaling for source pixels located close to a caustic. Indeed, first, if the QSO itself is close to the caustic (as it is the case for J1131), the residuals from the PSF subtractions mapped into the source plane will spoil the highly magnified central parts of the reconstructed host. Second but most important, the lensing model is never perfect, so that small but resolved photometric lensed structures in the image plane will not be mapped exactly at the same place in the source plane. However, if the pixel size in the source plane is not too small the lensed images will be mapped onto the same source pixel and the source substructures will even be enhanced, at the risk of being undersampled. It should also be noted that the S/N per source pixel is higher when its size is larger,  since a larger number of image pixels is mapped on that source pixel. We made several tests and checked that no gain in resolution was obtained with a source pixel scale smaller than half the image pixel scale.

\begin{table*}[t]
\begin{center}
\caption{Relative astrometry, flux ratio (with respect to image A) and absolute  photometry in filters F555W, F814W and F160W of the lensed QSO images A, B, C, D, as derived from PSF fitting (see Sect. \ref{sec:fit2}). After the Einstein ring has been subtracted, relative astrometry and absolute photometry are derived  for the lensing galaxy G from the best fit of the S\'ersic model (see Sec.\ref{sec:fit4}). After subtracting the lens, the astrometry of component X is derived from a gaussian fitting and its magnitude from aperture photometry (corrected for infinite aperture). Theoretical magnification ratios $M_{\rm A}/M_{i}$ and the (unobserved) AGN position are derived from our fiducial lens model (see Sect. \ref{sec:fit5}). Magnitudes of the host are calculated in Sect. \ref{sec:fit6}.   Quoted errors are standard errors on the mean, computed from the results on the NExp individual frames, including the systematics estimated in Sects. \ref{sec:red} and \ref{sec:photcal}.  Boldface indicates the final, weighed average astrometry derived from all bands.}
\label{tab:psf}
\begin{tabular}{rcccccc}
\hline
Object  & Filter & $\Delta\alpha \cos \delta$ & $\Delta \delta$ & $I_{\rm A}/I_i$ & m & $M_{\rm A}/M_{i}$\\
\hline
Image A & F555W & $0.000 \pm 0.000$ & $0.000 \pm 0.000$ & $1.000 \pm 0.000$ &$17.74\pm0.02$ & 1.00\\
  & F814W & $0.000 \pm 0.000$ & $0.000 \pm 0.000$ & $1.000 \pm 0.000$ &$ 17.43\pm0.02$ & 1.00\\
  & F160W & $0.000 \pm 0.000$ & $0.000 \pm 0.000$ & $1.000 \pm 0.000$ & $15.76\pm 0.04$ & 1.00 \\
B & F555W & $+0.035 \pm 0.003$ &$+1.192	\pm 0.004$ & $1.105 \pm0.020$ & $17.85\pm0.02$ & 1.74 \\
  & F814W & $+0.029 \pm 0.005$	&$+1.186\pm 0.004$ & $1.085\pm0.018$ & $17.52\pm 0.04 $ & 1.74\\
  & F160W & $+0.030 \pm 0.004$ & $+1.187\pm 0.004$ & $1.30\pm 0.030$ & $16.05\pm 0.03$ & 1.74\\
  & {\bf Mean} & {\boldmath $+0.032 \pm 0.002$} & {\boldmath $+1.188 \pm 0.002$} & - & - &{\bf 1.74}\\
C & F555W & $-0.584 \pm 0.004$&$ -1.115\pm 0.004$ & $2.941\pm  0.044$& $18.91\pm 0.02$ & 2.10\\
  & F814W & $-0.591 \pm 0.005$	&$-1.123\pm 0.004$ & $2.764\pm0.086$ & $18.50\pm0.02 $ & 2.10 \\
  & F160W & $-0.595 \pm 0.005$ & $-1.122\pm 0.005$ & $2.62\pm 0.06$ & $16.80\pm0.04$ & 2.10 \\
  & {\bf Mean} & {\boldmath $-0.590 \pm 0.003$} & {\boldmath $-1.120 \pm 0.003$} & - & - & {\bf 2.10}\\
D & F555W & $-3.110 \pm 0.003 $&$ +0.885\pm 0.003$ &$10.500 \pm0.248$ & $20.27\pm 0.05$ &  20.4\\
  & F814W & $-3.117 \pm 0.006$ &$+0.879\pm 0.004$ & $10.655\pm 0.348 $ &$20.00\pm 0.04$ & 20.4\\
  & F160W & $-3.123 \pm 0.009$ &$+0.896\pm 0.009$ & $11.12\pm 0.52$ & $18.37\pm0.06$ & 20.4\\
  & {\bf Mean} & {\boldmath $-3.112 \pm 0.003$} & {\boldmath $+0.884 \pm 0.002$} & - & - & {\bf 20.4}\\
Lens G & F555W & $-2.014 \pm 0.003$ &$+0.612\pm 0.003$ & - &$19.58\pm0.12$ & -\\
  & F814W & $-2.029 \pm 0.011$ &$+0.607\pm 0.005$ & - &$17.61\pm0.06$ &-\\
  & F160W & $-2.019\pm0.004$ & $+0.606\pm0.005$ & - & $15.75\pm0.08$ &-\\
  & {\bf Mean} & {\boldmath $-2.016 \pm 0.002$} & {\boldmath $+0.610 \pm 0.002$} & - & - &-\\
X & F555W & $-1.931\pm 0.011$& $+1.145\pm 0.008$& - & $25.30\pm 0.30$ &-\\
  & F814W & $-1.930\pm 0.005$& $+1.140\pm 0.007$ & - &$23.29\pm0.06$ & -\\
  & F160W & $-1.930\pm0.005$ & $+1.137\pm 0.005$ & - & $21.15\pm0.06$ &-\\
  & {\bf Mean} & {\boldmath $-1.930 \pm 0.003$} & {\boldmath $+1.139 \pm 0.004$} & - & - &-\\
Host H & F555W & - & - & - &$21.53\pm0.63$ &-\\
  & F814W & - & - & - & $19.49\pm0.15$ &-\\
  & F160W & - & - & - &$17.27\pm0.03$ &-\\
AGN  & {\bf Mean} & {\boldmath $-1.501 \pm 0.02$} & {\boldmath $+0.427 \pm 0.02$} & - & - &-\\
\hline
\end{tabular}
\end{center}
\end{table*}

\section{Fitting the gravitational lens system}
\label{sec:fitting}

The lens system J1131, with its bright lens, its highly structured ring and its 4 superimposed PSFs is morphologically complex. {\it Simultaneously} fitting the PSFs over a bright and variable background, the photometric and  the deflecting parameters of the lens and recovering the real source shape represent an utopic task! Instead, a {\it sequential} approach has been followed. We identified the following steps:

\begin{enumerate}
\item
{\it Fitting of a preliminary lens model} (e.g. Singular Isothermal Sphere (SIS) + shear) by using only the approximate astrometry of the point-like images (derived from gaussian fitting).
\item
{\it Fitting of the PSF positions and intensities}. The bright background ring is taken into account by simultaneously fitting a {\it parametrized} source (e.g. a truncated exponential disk profile), lensed by the preliminary model. Precise relative astrometry and photometry of the point-like images may then be derived in each band and PSFs can be removed from the original individual frames. 
\item
{\it Ring subtraction.} Reconstruct the source plane from image plane areas located outside the critical line (see details in Sect. \ref{sec:source})  and stack the resulting frames to get the best S/N result in each band. Map the source back in the image plane to build the averaged rings and subtract them from the PSF-subtracted images obtained in step-2.
\item
{\it Fitting of the lensing galaxy} with a convolved photometric galactic model (e.g. S\'ersic profile) on images obtained in step-3. Subtract the fitted lens model from the {\it PSF subtracted} images derived in step-2 and stack them to create the best image of the lensed host galaxy in each band.
\item
{\it Fitting of the final lens model}. At this stage, constraints from extended lensed structures can be added to the final astrometric constraints. Several lens models can be compared.
\item
{\it Source reconstruction} in the different bands with  the best lens model.
\end{enumerate}

We now describe in more details each step defined above. Note that all minimizations have been made using the {\sc Minuit} algorithm developed at CERN (James 1994). 

\begin{table*}[t]
\begin{center}
\caption{Best values of the S\'ersic and de Vaucouleurs (dVc) profiles fitted to the lensing galaxy (see Sect. \ref{sec:fit4} for details). Quoted errors are the 1$\sigma$ dispersion of the results on the NExp individual frames. The number of d.o.f. is about 3120 for F555W and F814W frames, and about 1100 for F160 frames.}
\label{tab:lensphot}
\begin{tabular}{cccccccc}
\hline
Filter & Model  & $\mu_e(z)$ (mag/arcsec$^2$) & $q_s$ & $\varphi$(P.A.) & $R_e$ (kpc $h^{-1}_{50}$) & n & $\chi^2$/d.o.f \\
\hline
F555W  & S\'ersic  & $21.84\pm0.38$ & $0.96\pm0.01$ & $-22^\circ\pm20$  & $4.69\pm0.77$ & $2.19\pm 0.22$ & 2.53 \\
       & dVc  & $23.35\pm0.21$ & $0.96\pm0.02$ & $-60^\circ\pm32$  & $10.48\pm0.88$ & $4$ & 2.76 \\
F814W &S\'ersic  & $20.36\pm0.17$ & $0.94\pm0.01$ & $-56^\circ.1\pm2.6$ & $5.05\pm0.37$ & $2.67\pm 0.12$ & 3.62 \\
      & dVc & $21.49\pm0.05$ & $0.93\pm0.01$ & $-64^\circ.2\pm1.9$ & $9.33\pm0.18$ & $4$ & 3.98 \\
F160W & S\'ersic  & $18.09\pm0.22$ & $0.92\pm0.01$ & $-62^\circ.2\pm4.1$& $4.16\pm0.41$ & $2.71\pm 0.16$ & 18.8 \\
      & dVc  & $19.28\pm0.14$ & $0.91\pm0.01$ & $-65^\circ.3\pm3.8$& $8.33\pm0.51$ & $4$ & 19.9 \\
\hline
\end{tabular}
\end{center}
\end{table*}

\subsection{Fitting a lens model from astrometry}
\label{sec:fit1}

The astrometric positions of the point-like images yield powerful and robust constraints on the GL model. Indeed, unlike the flux ratios, they are not affected by external or internal pertubations such as differential extinction in the lensing galaxy (Nadeau et al. 1991\cite{nadeau91}, Jean and Surdej 1998\cite{jean98}), delayed intrinsic variations, microlensing (e.g. Wambsganss 2001\cite{wambsganss01}) or substructures present in the gravitational potential (e.g. Brada\v c et al. 2004\cite{brada04}).  Those photometric effects can easily reach tens of percent while the astrometry of the point-like images is only affected at the scale of the Einstein ring of the substructures, which is of the order of or smaller than the astrometric measurement errors.

The weakness of the astrometric constraints is their scarcity. After deducting 4 degrees of freedom for the lens and source positions, only 6 parameters can be fitted from the astrometric constraints offered by J1131. This is however more than enough to determine the strength (i.e. the Einstein radius $\theta_E$)  and  the azimutal anisotropy (either internal [ellipticity $e$ and orientation $\varphi_e$] or external [shear $\gamma$ and orientation $\varphi_\gamma$]) of the lens potential.

In order to avoid problems associated with the computation of errors in the source plane close to the caustic, we minimized in the image plane the quantity:

\begin{equation}
\chi^2_a = \left(\frac{x_g - x_l}{\sigma_{x_g}}\right)^2 + \left(\frac{y_g - y_l}{\sigma_{y_g}}\right)^2 + \sum_{k=1}^{4} \left[\left(\frac{x_{k}-\hat{x}_k}{\sigma_{x,k}}\right)^2 + \left(\frac{y_{k}-\hat{y}_k}{\sigma_{y,k}}\right)^2\right]\,,
\label{eq:chi2a}
\end{equation}

\noindent
where $(x_g,y_g)$ and $(x_k,y_k)$ are the observed positions of the lensing galaxy and the $k^{\rm th}$ point-like image, with associated errors $(\sigma_{x_g},\sigma_{y_g})$ and $(\sigma_{x_k},\sigma_{y_k})$, and $(x_l,y_l)$ and $(\hat{x}_k,\hat{y}_k)$ are the fitted lens and image positions. 

Following Sluse et al (2003\cite{sluse03}), we fitted a simple  SIS + $\gamma$ model for this preliminary step.

\subsection{Fitting and subtracting the PSFs}
\label{sec:fit2}

In order to reconstruct the host galaxy in the source plane, the PSFs must first be removed. However the PSF flux determinations are contaminated by the bright Einstein ring (and conversely), especially in the F814W and F160W filters (see discussion in paper I). To minimize this problem, the ring is simultaneously modeled as the image of a truncated exponential profile, lensed by the SIS + $\gamma$ model. The parameters of the source are simultaneously fitted with the PSF intensities and positions by minimizing:

\begin{equation}
\chi^2_e = \sum_{i,j} \left(\frac{I_{ij}-\hat{I}_{ij}}{\sigma_{ij}}\right)^2,
\label{eq:chi2e}
\end{equation}

\noindent
where $I_{ij}$ is the observed intensity in pixel (i,j) with error $\sigma_{ij}$ and $\hat{I}_{ij}$ is the sum of the intensities of the modeled ring and of each PSF in pixel $(i,j)$. $\chi^2_e$ is computed in a small region including the PSFs and the inner part of the ring. Note that the lens model is not used to determine the PSF positions but only the shape of the ring.

The initial conditions are provided by the PSF positions given by simple gaussian fittings of the point-like images, by an extended circular source and by the SIS + $\gamma$ fitted in Sect. \ref{sec:fit1}. After each individual frame is processed, the  astrometry and flux ratios relative to image A and the magnitudes of all point-like images  are averaged on NExp separately in each spectral band. The results are reported in Table \ref{tab:psf}. The quoted errors are $1 \sigma$ standard errors (i.e. 68.3\%) on the mean.

The derived astrometric results of the four lensed QSO images are mutually consistent between each photometric band. This is reassuring and allows us to compute the final weighed averaged astrometry also listed in Table \ref{tab:psf} in bold characters.

\subsection{The lens and the X component}
\label{sec:fit4}

Once the PSFs have been subtracted, most of the light associated with the Einstein ring can then be removed as explained in {\it step-3}. 

Eventually, only the lensing galaxy G and component X are remaining on the frames and the fit of the lens can be made without being heavily biased by the ring contribution. Of course, it is still necessary to mask out the regions where ring residuals are seen, but X is not thought to bias the lens fitting since it is very compact, well separated and much fainter than the lens.

We made use of a  S\'ersic profile (S\'ersic 1968\cite{sersic68}) to describe the surface brightness $\Sigma$ of the lens:

\begin{equation}
\Sigma_{\rm Sersic} = \Sigma_e \rm{e}^{-k\left[\left(\frac{r}{r_e}\right)^{1/n}-1\right]}\,,
\label{eq:sersic}
\end{equation}

\noindent
where $r_e$ is the effective radius including half of the total flux, $\Sigma_e$ is the surface brightness at $r_e$ and $n$ is the power law index ($n=4$ corresponds to a pure de Vaucouleurs profile prototyping elliptical galaxies,  while $n=1$ is equivalent to an exponential profile describing galactic disk profiles) and $k$ is a normalization coefficient: $k=1.9992 n -0.3271$ (e.g. Simard et al. 2002\cite{simard02}). Because the isophotes can be elliptical, $r$ is parametrized as $r^2(x,y) = x^2 + y^2/q_s^2$ where $q_s$ is the axis ratio of the S\'ersic model. $Oxy$ is the referential attached to the lens principal axes, whose positional angle is $\varphi$.  Before fitting, the S\'ersic profile was convolved with the appropriate Tiny Tim PSF (see Sect. \ref{sec:red}). 

Minimizing equation \ref{eq:chi2e} with $\hat{I}_{ij}$ given by the latter model yielded the best parameter values. We also used the {\sc Galfit} fitting algorithm (Peng et al. 2002\cite{peng02}) to check the robustness of our results. 

The best model position relative to image A and the best model total magnitude are reported for each band in Table  \ref{tab:psf}. The best shape parameters are displayed in Table \ref{tab:lensphot}, where $\mu_e(z) = -2.5\log \Sigma_e/${\it Scale}$^2$ is the standard surface magnitude at $r_e$ ({\it Scale} is the pixel scale on the sky) and the physical size of the effective radius $R_e = D_{A,l} r_e$, where $D_{A,l}$ is the angular cosmological distance to the lens $l$. We provide not only the shape parameters of the best S\'ersic model, but also the ones corresponding to the best de Vaucouleurs model (i.e. when keeping $n=4$ constant), for further analysis and comparison with previous studies of the fundamental plane (see Sect. \ref{sec:fp}).

 The  main contribution to the $\chi^2$ value (see Table \ref{tab:lensphot}) is a faint, spatially variable, {\it diffuse background} remaining within the Einstein ring. This background is very likely associated with secondary images of faint, remote, off-axis regions of the extended host galaxy. This effect is wavelength dependent since the host galaxy is  much brighter and extended in the NIR (see Sect. \ref{sec:fit6}).  

\begin{figure}
\hbox to \columnwidth{
{\includegraphics[width=2.5cm]{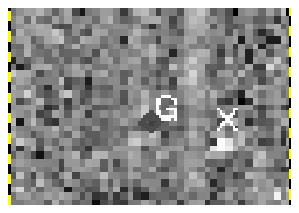}}
{\includegraphics[width=2.5cm]{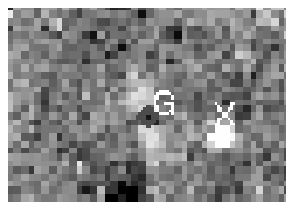}}
{\includegraphics[width=2.5cm]{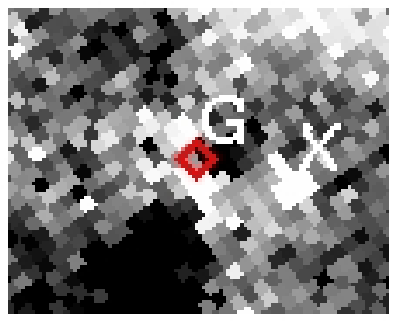}}
}
\caption{Residuals of the central $2''\times1.5''$, in F555W (left), F814W (center) and F160W (right). The grey scale is in the range $-5\sigma,+5\sigma$ (orientation of the F555W images).}
\label{fig:residuXG}
\end{figure}

On the other hand, {\it extended} residuals are seen close to the lens centroid, in the F814W and F160W bands, at the 4$\sigma$ and 7$\sigma$ level respectively (for comparison, the X object leaves residuals at the 15 and 27 $\sigma$ level, respectively; see Fig. \ref{fig:residuXG}). Although the significance is a little bit less in the F814W band, the higher angular resolution and the lower background clearly reveal  a double-object-like signature, more or less  oriented along the main axis of the fitted S\'ersic model. 

We made several tests to identify the possible origin of those residuals, but without much success. First, they are definitely {\it not} associated with a putative fifth lensed image, since a S\'ersic-plus-point-like model does not improve the fit. Second, a simple bulge+disk decomposition does not solve the problem, even whitout co-centering. Third, allowing for boxy isophotes (e.g. Peng et al. 2002\cite{peng02}) does not change the shape of the residuals. Eventually, we simultaneously fit two S\'ersic profiles {\it restricted} to the most inner part of the lens ($0.75''$ diameter). The central residuals are then removed. However,  a larger $\chi^2$ is obtained in the wings, the S\'ersic exponents are very low ($n_1=1.58$ and $n_2 = 0.82$), and most importantly, the results cannot be reproduced in the F160W band. Indeed, fixing the relative position of the two components to those obtained in the F814W band leads to unphysical values of the parameters in the F160W band without improving the $\chi^2$ value. 

We also investigated possible artefact origins. For example, the ACS distortion correction and the poorly sampled PSF might introduce artefacts during resampling. But the results are stable from frame to frame obtained at different dithered positions.  We recently found out that galaxy-core residuals in S\'ersic-subtracted F814W data have also been noticed by Bolton et al. (2006)\cite{bolton06} who get rid of them by applying a more complex B-spline galaxy model.

Finally, we checked that the flux associated with the residuals is smaller than 1\% of the total flux of the lens. Lacking high angular resolution 2-D spectroscopy, we decided to keep the results of the one-component fitting analysis, even if the spatial extension of the residuals in F814W (roughly along the Oy CCD axis) is probably responsible for the slight offset and the larger error of the relative position of the lens in right ascension, with respect to the other bands (see Table \ref{tab:psf}). Differential imaging performed after renormalization between the different bands confirms the presence of a slight offset of the lens in the F814W band.  However a map of the F555W-F160W apparent colour index does not show any fluctuation above the noise (0.12 mag rms)  which could account for the presence of a dust lane.

A more detailed discussion and analysis of the lens properties is postponed to Sect. \ref{sec:lightmass}.

\vspace{0.25cm}
The lens photometric model being subtracted, the last component seen on the residuals is the compact, unresolved X-object already identified in Paper I on the NICMOS frames. It is also seen on the F555W and F814W ACS images. In each band, its position relative to image A has been derived from gaussian fits and its magnitude has been obtained from flux measurements within a 2 pixel radius aperture corrected for infinite aperture. Those positions and magnitudes are listed in Table \ref{tab:psf}. The discussion about the nature of X is delayed to Sect. \ref{sec:X}.

\begin{table*}[t]
\begin{center}
\caption{Fitted values of the parameters of simple  mass models (see text for details). For each model, the first (resp. second) line gives the results when fitting (resp. not fitting) the lens position (LF; resp. NLF). $\theta_E$ is the angular Einstein radius, $\varepsilon$ the intrinsic ellipticity, $\gamma$ the external shear, $\theta_\varepsilon$ and $\theta_\gamma$ the position angles, $(\Delta_\alpha,\Delta_\delta)$ the differences between the modeled and the observed lens positions in $\alpha$ and in $\delta$.}
\label{tab:lensmod1}
\begin{tabular}{rc|ccccccccccc}
\hline
Model &  & $\theta_E('')$ & $\varepsilon$ & $\theta_\varepsilon$(P.A.) & $\gamma$ & $\theta_\gamma$(P.A.) & $\Delta_\alpha('')$ & $\Delta_\delta ('')$&$\chi_a^2$/$d.o.f._a$ & $\chi_e^2$/$d.o.f._e$ &d.o.f.$_a$ & d.o.f.$_e$\\
\hline
SIS + $\gamma$ &LF & 1.842& $-$ & $-$  & 0.145 & -73.9 & 0.021 & -0.052 & 190 & 42 & 3 &-  \\
               &NLF& 1.842 & $-$ & $-$ & 0.144 & -74.3 & 0.023 & -0.068 & 183 &  44 & 1 & 674\\
SIE & LF & 1.823 & 0.45 & -73.6 & $-$ & $-$ & -0.015 & -0.041 & 272 & 48 & 3 & -\\
    & NLF & 1.826 & 0.48 & -74.2 & $-$ &$-$ & -0.059 & -0.045 & 418 & 39 & 1 & 674\\
SIE + $\gamma$   & LF & 1.849 & 0.11 & -49.9 & 0.130 & -80.6 & 0.000 & -0.034 & 177 & 38 & 1 & -\\ 
   & NLF & 1.850 & 0.11 & -50.1 & 0.130 & -80.6 & 0.005 & -0.044 & 10 & 39 & 1 & 672\\ 
\hline
\end{tabular}
\end{center}
\end{table*}

\begin{table*}[t]
\begin{center}
\caption{Fitted values of the parameters of best SIE + $\gamma$ model after introducing an $m=4$ multipole ($\varepsilon_4, \theta_4$). The lens position is fitted. Same notations as in Table \ref{tab:lensmod1}.}
\label{tab:multipole}
\begin{tabular}{ccccccccccccc}
\hline
$\theta_E('')$ & $\varepsilon$ & $\theta_\varepsilon$(P.A.) & $\gamma$ & $\theta_\gamma$(P.A.) & $\varepsilon_4('')$ & $\theta_4$(P.A.) & $\Delta_\alpha('')$ & $\Delta_\delta ('')$&$\chi_a^2$/$d.o.f._a$ & $\chi_e^2$/$d.o.f._e$ &d.o.f.$_a$ & d.o.f.$_e$\\
\hline
1.848 & 0.177 & -60.78 & 0.124 & -85.01 & -0.03 & 11.05 & $<$0.001 & $<$0.001 & 21.9 & 38.9 & 1 & 670 \\
\hline
\end{tabular}
\end{center}
\end{table*}

\subsection{Fitting different lens models}
\label{sec:fit5}

In order to determine the optimal lens model, we used  the  best available astrometry  with a conservative positional error of 3 mas for each component (see Sect. \ref{sec:fit2} and Table \ref{tab:psf}). On the other hand, making use of the lens photometric model determined in Sect. \ref{sec:fit4}, we built PSF-subtracted and lens-subtracted images only displaying the lensed host galaxy. The latter images show many {\it extended sub-}structures which may be used as new constraints on the lens model (see discussion in Sect. \ref{sec:intro}).

 Those constraints have been combined to the astrometric constraints in the minimisation process through the relations (\ref{eq:chi2a}) and (\ref{eq:chi2e}), according to the following relation:

\begin{equation}
\chi^2 = \lambda \chi^2_a/{\rm d.o.f}_a + (1-\lambda)\chi^2_e/{\rm d.o.f}_e\,,
\label{eq:chi2}
\end{equation}

\noindent
where $\lambda$ is a weighing factor, d.o.f.$_{a,e}$  are the numbers of degrees of freedom associated with the astrometry and the extended image respectively. The value of d.o.f.$_e$ is obtained by subtracting the number of parameters from the number of pixels, but for the astrometric constraints, we fixed the minimum value of d.o.f.$_a$ to 1. Should the number of parameters be too large for the number of astrometric constraints, the d.o.f.$_e$ is more than sufficient.

For each step of the fit, $\chi^2_e$ was computed from the reconstructed ring intensities  $\hat{I}_{ij}$. The sum can only be made over pixels $(i,j)$ which are multiply imaged (otherwise $I_{ij}-\hat{I}_{ij}$ is identically zero)\footnote{Strictly speaking, the $\chi^2$ estimator cannot be applied since the computed pixel values are not independent from the observed ones. However, the obtained values for the $\chi^2$  being not acceptable from a pure statistical point of view, a more appropriate statistical test is not critical.}. More specifically, the sum was restricted  over 681 pixels belonging to the {\it sub-structures} of the ring well visible in the F555W band.

\subsubsection{Canonical lens models}
\label{sec:canonical}

First, we dealt with three simple models, namely the Singular Isothermal Sphere plus external shear (SIS$+\gamma$, e.g. Schneider et al. 1992\cite{schneider92}, Claeskens et al. 2001\cite{claes01}), the Singular Isothermal Ellipsoid (SIE) (Kassiola and Kovner 1993\cite{kassiola93}, Kormann et al. 1994\cite{kormann94}, Keeton and Kochanek 1998\cite{keeton98}) and the SIE$+\gamma$. We adopted the SIE normalization proposed by Kormann et al. (1994\cite{kormann94}), where the projected mass within a given surface density contour is independent of the axis ratio $q$ of the matter distribution. The ellipticity $\varepsilon$ is related to $q$ through $q = \sqrt{\frac{1-\varepsilon}{1+\varepsilon}}$ (e.g. Keeton and Kochanek 1998\cite{keeton98}). 

Results are given in Table \ref{tab:lensmod1}, firstly when only fitting the astrometry of the 4 point-like images and of the lens (i.e. $\lambda=1$ in Eq. \ref{eq:chi2}; LF case). The dominant contribution to the $\chi_a^2$ comes from the lens position, which is not well reproduced by any of the models. So, when we added new constraints by  including extended structures ($\lambda=0.9$), we removed the lens position from the fit, hence the ``no lens fit'' (NLF) case in Table \ref{tab:lensmod1}.

 A striking result is that in the LF cases the astrometric positions are better reproduced in term of $\chi_a^2$ when an external shear is present (i.e. with the SIS + $\gamma$ or the SIE + $\gamma$ model). However, when the lens position is not fitted (NLF cases), the astrometry of the point-like images is much better reproduced with the SIE + $\gamma$ model ($\chi^2_e=10$),  as generally expected when both internal and external sources of anisotropy are included in the model (Keeton et al. 1997\cite{keeton97}).   However, the offset between the predicted and observed lens positions is significant since it amounts to about  14 $\sigma$. 

The SIE + $\gamma$ model is also slightly favoured by the extended structures, in terms of $\chi^2_e$. So despite the lens position offset, we shall call this latter model the fiducial lens model in the following.

Let us come back to the constraints offered by the extended substructures. The $\chi_e^2$/d.o.f.$_e$ values listed in Table \ref{tab:lensmod1} are not {\it strongly} sensitive to the adopted lens model! Anticipating on Fig. \ref{fig:ident} where corresponding substructures are correctly identified with our fiducial model -- meaning a reasonably good {\it astrometric} matching --, we can check on the direct F555W and F814W images that the surface brightness is  {\it not} identical between the corresponding substructures. The hypothesis of surface brightness conservation is thus {\it violated} . So, even a perfect lens model would lead to a significant disagreement in terms of $\chi^2_e$. The reason why the surface brightness is not preserved may be manifold. First, in the case of the present HST data of J1131, the thin arcs are probably not resolved radially and could thus be partially amplified. Second, spatially resolved sources still usually consist of many unresolved sources on which microlensing might act, resulting in a possible local violation of the surface brightness conservation hypothesis. This would especially be true at high angular resolution. Third, differential extinction in the lensing galaxy may also affect the apparent surface brightness, although, it should not be very efficient in the F814W band. All those arguments should constitute a warning for lens modeling  only relying on extended structures.

In this context, the $\gamma - \varepsilon$ degeneracy is not entirely raised by the extended structures. However the asymmetric cusp configuration of J1131 is more constraining than symmetric quads (see the case of \object{B1422+231} in Keeton et al. 1997\cite{keeton97}).

Finally, let us mention that the source position and theoretical amplification ratios of the point-like images given by our fiducial model are listed in Table \ref{tab:psf}. As already pointed out by Sluse et al. (2003\cite{sluse03}), they do not agree with the observed flux ratios. Microlensing is indeed observed in at least one component, but substructures in the lens potential do not seem to be favoured (see Paper I for a detailed discussion of this issue).

\subsubsection{Influence of the X-object}
\label{sec:Xlens}

As demonstrated in Sect. \ref{sec:X}, the X-object is very likely associated with the lens. Thus, one can wonder whether it may influence the deflection potential of the main lens and whether it is responsible for the reported offset between the predicted mass centroid and the observed light centroid of the lens (see Table \ref{tab:lensphot}). Regarding the second question, the X-object can not be the cause of the offset, since it is located on the North side of the lens (see Fig. \ref{fig:rgb} and Table \ref{tab:psf}), while the reported offset is essentially in the South direction.

To answer the first question, we followed two different approaches. In the first one, using the {\it lensmodel v1.06} software (Keeton 2001\cite{keeton01}), we modeled the X-object as an SIS, with its  position fixed to the observed one. Adding this component to the SIS + $\gamma$ model did not improve the $\chi^2_a$ and the value of the Einstein radius was found to be extremely small ($\theta_{\rm E,X} \sim 10^{-5}$$''$). In the second approach, we fitted several non-parametric asymmetric models (searched by a Monte-Carlo method), using the {\sc PixeLens} public software (Saha \& Williams 2004\cite{saha04}). We failed to predict any mass overdensity in the vicinity of X. Thus, both results tend to show that the X-object does not play an important role in the lens potential.

\subsubsection{More general lens models}
\label{sec:generalmod}

Although the number of astrometric constraints is limited and despite the fact that the extended substructures may not be very helpful because  the hypothesis of conservation of the surface brightness may be violated  (see Sect. \ref{sec:canonical}), it is tempting to use more complex lens models in order to try to better reproduce the observed lens position. We have tested the Softened Power Law Elliptical Mass Distribution (SPEMD) and the addition of higher order multipoles to the SIE + $\gamma$ model.

The deflection angle of the SPEMD can be efficiently computed with the {\sc fastell} algorithm provided by Barkana (1998)\cite{barkana98}. Allowing a variable index of the radial profile and/or a finite core radius and proceeding as in Sect. \ref{sec:canonical} did not improve the fit, which indeed converged toward the singular, isothermal profile.

 The lens potential may be written in terms of a multipole expansion (Kochanek 1991\cite{koch91}, Schneider et al. 1992\cite{schneider92}). While the monopole ($m=0$) is related to the radial profile and the quadrupole ($m=2$) is arising from an elongated mass distribution, the octupole term ($m=4$) indicates the presence of quadrangularity (e.g. boxiness/diskiness; Trotter et al. 2000\cite{trotter00}), which is observed in the isophotes of some elliptical galaxies (e.g. Bender et al. 1989\cite{bender89}, Rest et al. 2001\cite{rest01}). Such a term in the lens potential is also known to influence the magnifications of the point-like lensed images and has been introduced in the context of ``anomalous'' observed flux ratios (Evans \& Witt 2003\cite{evans03}, Keeton et al. 2003\cite{keeton03}). 

 Since the observed flux ratio of J1131 are not well reproduced by quadrupolar models (see Paper I and Table \ref{tab:psf}), we had one more reason to incorporate higher order terms in the lens potential. The {\it lensmoded v1.06} software (Keeton 2001\cite{keeton01}) has been used to simultaenously fit the astrometry of the point-like images and of the lens, as well as the observed flux ratio $I_{\rm B}/I_{C}$, which is  thought not to be affected by microlensing or by differential reddening (see Paper I). We found that {\it both} issues of the ``anomalous'' flux ratio and of the lens astrometry could be solved with an octupole (but not with an $m=3$ term). This result was confirmed by introducing an octupole in our modelisation following the definition (with $m=4$) :

\begin{equation}
\phi_m = - \frac{\varepsilon_m}{m}\cos m(\theta - \theta_m)\,
\label{eq:multi1}
\end{equation}

\noindent
and performing a new fitting. The values of the parameters of the best SIE + $\gamma$ + octupole model are given in Table \ref{tab:multipole}. Note that the amplitude of the perturbation $\varepsilon_4$ can be related to the amplitude $a_4$ of the deviation of the isophotes of the surface mass density, assuming that the unperturbed profile is isothermal:

\begin{equation}
a_m = \frac{\varepsilon_m (m^2-1)}{m \theta_E}\,.
\label{eq:multi2}
\end{equation}

The discussion of this model is made in the context of Sect. \ref{sec:lightmass}.

\begin{figure}
\begin{center}
\includegraphics[width=\columnwidth]{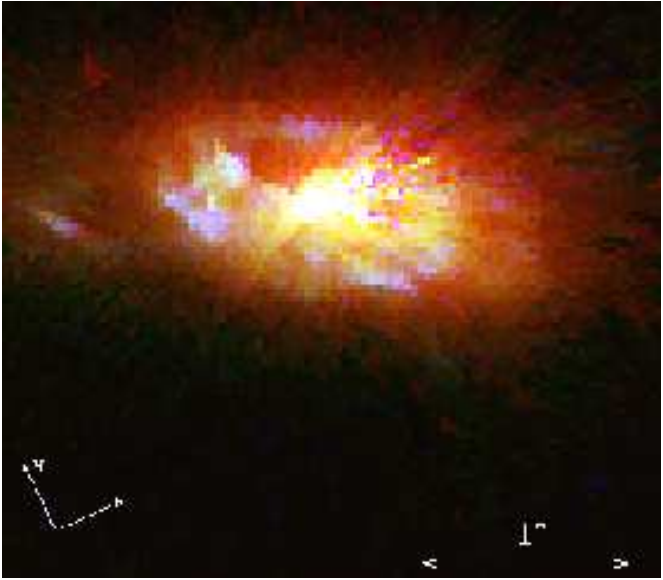}
\caption{Colour image of the reconstructed host galaxy of RXS J1131-1231, combining the ACS F555W (blue), ACS F814W (green) and NICMOS2 F160W (red) images.}
\label{fig:source_rgb}
\end{center}
\end{figure}

\begin{table*}[t]
\begin{center}
\caption{Best values of parameters of the S\'ersic profile fitted on the reconstructed host galaxy.}
\label{tab:host}
\begin{tabular}{cccccccc}
\hline
 Filter & $\mu_e(z)$ (mag/arcsec$^2$) & $q_s$ & $\varphi$(P.A.) & $R_e$ (kpc $h^{-1}_{50}$) & n & $m_{\rm Sersic, tot}$ \\
\hline
F160W & 18.53 & 0.49 & -67.8& 9.8 & 1.05 & 17.16\\
F814W & 21.15 & 0.48 & -66.5& 14.4 & 1.25 & 19.01\\
\hline
\end{tabular}
\end{center}
\end{table*}

\subsection{The host galaxy}
\label{sec:fit6}

The host galaxy is finally reconstructed from the median of the stacked PSF-subtrated and lens-subtracted ACS and NICMOS images (see Sect. \ref{sec:fit2} and \ref{sec:fit4}), as described in Sect. \ref{sec:source}. The adopted lens model is the best SIE + $\gamma$ + octupole described in Sect. \ref{sec:generalmod}. The multicolour reconstructed image of the host in the source plane is shown in Fig. \ref{fig:source_rgb}. Let us already note here that adopting instead the SIE + $\gamma$ fiducial lens model defined in Sect. \ref{sec:canonical} yields qualitatively the same results.

In order to avoid artifact close to the center, we made aperture photometry in a ring with inner and outer radii of $0.12''$ and $1.5''$ respectively. Since the source is reconstructed in each photometric band from the unique median image, it is impossible to derive error bars from the dispersion of the results. Rather, we computed the Einstein ring corresponding to the reconstructed source (whose regions with $r<0.12''$ and $r>1.5''$ have been masked) and we subtracted it from each image in the image plane. We derived the error on the magnitude from the dispersion of the residuals. This technique overestimates the noise since many pixels in the source plane have an increased S/N thanks to multiple imaging. Photometric results are listed in Table \ref{tab:psf}, as well as the position of the central AGN predicted by our fiducial model. The large astrometric error bars reflect the uncertainty on the predicted lens position.

On the other hand, the ratio between the flux of the Einstein ring built as described above and the flux of the reconstructed source gives the total magnification of the host in each band. We found $M_{\rm 555} = 10.9$,  $M_{\rm 814} = 9.9$ and $M_{\rm 160} = 7.8$. The decrease of the magnification from F555W to F160W corresponds to the fact that the source is more extended in the near IR, well beyond the caustic.

Finally, although the host is quite irregular, we tried to fit a S\'ersic profile to quantify its morphology. We kept the host position fixed to the AGN position computed by the lens model and we considered a thick ring with $0.12'' < r < 1.5''$. Since the noise in the source plane is ill-defined, the  $\chi^2$ value and formal errors are meaningless. We preferentially used the F160W image where the host profile is the smoothest, but we checked the stability of the results by also fitting the F814W images. Results are listed in Table \ref{tab:host}. Structural parameters  look rather stable, although the effective radius and the slope are slightly modified in the F814W band. The host galaxy profile is best reproduced by a S\'ersic model with $n\sim 1$ typical  of {\it disk dominated} galaxies (e.g. de Jong et al. 2004\cite{dejong04}).   Although the host is globally well subtracted, positive residuals are correlated with emission in the F555W filter, while negative residuals  are seen around the center and might indicate that the PSFs have been slightly oversubtracted. Total magnitudes  of the fitted model in a box of $4''\times4''$ are also given in Table \ref{tab:host}. Again, analysing the host galaxy reconstructed with the fiducial model yields results very close to those presented in Table \ref{tab:host}.

The physical properties of the host are discussed in Sect. \ref{sec:host}.

\begin{figure}
\begin{center}
\includegraphics[width=\columnwidth]{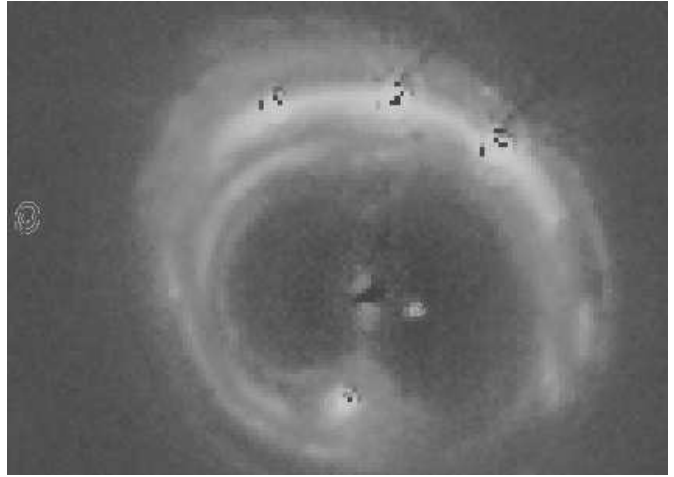}
\caption{Search for the primary image corresponding to X with the fiducial lens model: no detection around the left calculated contours.}
\label{fig:identX}
\end{center}
\end{figure}

\section{Results and discussion}
\label{sec:result}

\subsection{The X-object and the lens}
\label{sec:reslens}

\subsubsection{The X-object}
\label{sec:X}

First, although X does not seem to be resolved, it cannot be the fifth lensed image of the central AGN. Indeed, besides the fact that its colours are completely different from the ones of images A-D (see Table\ref{tab:psf} and Fig. \ref{fig:rgb}), a non-singular lens model would only predict a faint lensed image roughly aligned with G and D.

Second, X is not a secondary image of a substructure in the source. If such was the case, the primary image should have been detected since it would be well separated from the ring and at least as bright as X, depending on whether it is resolved or not. Fig. \ref{fig:identX} shows that it is not the case. 

Third, X is not a foreground star.  Indeed, Table \ref{tab:psf} shows that its apparent colours are (F555-F814)$_X = 2.01\pm0.31$ and (F814-F160)$_X = 2.14\pm0.08$. Stars with $V-I$ $\simeq$ 2 are either giant K5 or M0 dwarfs (Drilling and Landolt 1999\cite{drilling99}). But these stars have V-H smaller than 3.6 (Binney and Merrifield 1998\cite{binney98}) and are consequently bluer than the V-H colour index observed for X.

In conclusion, X is very likely  a satellite dwarf elliptical galaxy of the lens. Indeed, its apparent colours are very similar to that of the lens ((F555-F814)$_G = 1.97\pm0.14$ and (F814-F160)$_G = 1.86\pm0.10$; see Fig. \ref{fig:rgb}). Its restframe colours ($(B-r)_z = 1.31\pm0.31$ and $(B-J)_z = 3.50\pm0.31$) are comparable to those of local ellipticals (see Sect. \ref{sec:fp}).

\subsubsection{Comparing light and mass distributions}
\label{sec:lightmass}

As already noted by several authors, the pure $1/r^4$ de Vaucouleurs law does not correspond to the best fitting of the surface brightness profile of ellipticals. Although the values we obtain for the slope parameter are significantly smaller than the canonical $n=4$ value, they remain typical of early-type galaxies, which exhibit quite a broad range of values (e.g. de Jong et al. 2004\cite{dejong04}, di Serego Alighieri et al. 2005\cite{alighieri05}).

The structural parameters of the best S\'ersic model of the lens are reported for each band in Table \ref{tab:lensphot}. Its shape is pretty circular, especially in the F555W filter, where the derived orientation is meaningless. In the red bands, the averaged axis ratio is $q_s = 0.936\pm0.004$  and is close to the axis ratio of the mass distribution, $q_m=0.84$ -- $0.91$, as derived from the fiducial and best models in Tabs. \ref{tab:lensmod1} and \ref{tab:multipole}. Similarly, the averaged orientation of the light distribution (P.A = $-57.8\pm2.2$ degree) is comparable  to the orientation of the mass model ($ -50.1 < $ P.A. $< -60.8$ degrees).  This seems to indicate a good alignment between the projected luminous and dark matter distributions.

On the other hand, we noticed in Sect. \ref{sec:canonical} that the mass centroid of  the fiducial model had to be offset by about $0.04''$ with respect to the photometric centroid to reproduce the image positions, unless an octupole is added to the lens model (Sect. \ref{sec:generalmod}). The presence of an octupole might be related to the fact that photometric residuals are also seen in the F814W band after subtracting the best S\'ersic profile. It is also striking that the octupole simultaneously helps in reproducing the observed flux ratio $I_{\rm B}/I_{C}$. 

However the physical origin of this octupole is difficult to identify. Indeed, to improve the fits, the octupole position angle must be completely different from that of the quadrupole. This means that the solution ($\varepsilon_4, \theta_4$) is degenerated with the solution (-$\varepsilon_4, \theta_4+45^\circ$) and that the boxiness/diskiness cannot be assessed. Using Eq. \ref{eq:multi2}, the surface mass density isophotes are characterized by  $|a_4| = 0.06$. This value is quite high when compared to the boxiness/diskiness of light isophotes observed in elliptical galaxies (Rest et al. 2001\cite{rest01}). Besides that, as we saw in Sect. \ref{sec:fit4}, allowing for the boxiness of the isophotes did not remove the residuals observed in the F814W band. And why are those residuals mainly seen in the red bands? The mass octupole might thus be a (non-unique) mathematical way of accounting for some pertubations in the potential, for example due to interactions with a satellite galaxy (X-object?).   This puzzling situation could be investigated with 2-D spectroscopy obtained at high angular resolution.

In conclusion, while low-mass halo substructures do probably not provide the correct explanation for the observed 'anomalous' flux ratios of J1131 (see paper I), the $m=4$ multipole can reproduce the observed flux ratios and lens position but its physical nature is difficult to identify, as it is for other lenses (Kochanek \& Dalal 2004\cite{koch04}).

Finally, note that the value of the external shear (see Table \ref{tab:multipole}) is also found to be quite high. However its direction points towards the position (affected by large error bars) of a group of galaxies identified at the lens redshift from a red sequence analysis by Williams et al. (2006\cite{williams06}).

\subsubsection{Lens colours and evolution}
\label{sec:fp}

Despite the uncertainties relative to the lens model described in the previous section, the monopole term, i.e. the value of the Einstein ring radius, remains quite robust. This makes possible a deeper investigation of the lens.

In order to seek for galaxy evolution, the lens must be located with respect to the {\it local} FP, defined for example by the Coma cluster FP. Following the analysis by van de Ven et al. (2003\cite{vandeven03}) based on previous works by J\o rgensen, Franx and Kj\ae rgaard (1995a\cite{jorg95a}, 1995b\cite{jorg95b}, 1996\cite{jorg96}), the Coma FP relation can be written as:

\begin{equation}
\log R_e = \alpha \log \sigma_c + \beta \mu_e + \gamma\,,
\label{eq:fp}
\end{equation}

\noindent
where $\alpha=1.24\pm0.07$, $\beta = 0.328 \pm 0.001$ and $\gamma = -9.12\pm 0.05$ in the Gunn r filter. In Eq. (\ref{eq:fp}), $R_e$ is the effective radius in kpc, $\sigma_c$ is the central velocity dispersion in km/s and $\mu_e$ is the surface brightness at $r_e$ in mag arcsec$^{-2}$. 

 In a given band, the evolution of the M/L ratio with respect to local galaxies can be quantified  from the FP analysis using the following relation (Treu and Koopmans 2004\cite{treu04} and references therein):

\begin{equation}
\Delta \log (M/L) = -0.4\frac{\Delta \gamma}{\beta}\,,
\label{eq:ml}
\end{equation}

\noindent
 where $\Delta \gamma = \gamma_{\rm obs} - \gamma$ . This relation is independent of the possible evolution of the slopes $\alpha$ and $\beta$ (Treu et al. 2005b\cite{treu05b}).

 Since one easily shows that overestimating $\mu_e$ or $\sigma_c$ in Eq. \ref{eq:fp} would hide the lens evolution, we first discuss the determination of those parameters.

\vspace{0.5cm}
\noindent
{\large {\it i) $\mu_e$ and the lens colours}}
\vspace{0.25cm}

Since the  B, (Gunn-)r and J bands redshifted at $z=0.295$ nearly correspond to the observed ones F555W, F814W and F160, respectively, it is easy to derive the $(B-r)_z$ and $(B-J)_z$ {\it rest frame} lens colour indices. It is also possible to estimate the {\it rest frame} surface brightness in the r-band $\mu_{e,r}(z=0)$ needed to locate the lens with respect to the local FP. In the latter case, we used the following colour transformation:

\begin{eqnarray}
\mu_{e,r}(z=0)& = &\mu_{e,F814W}(z)  + (r_{z,{\rm AB}} - F814W_{\rm AB}) - (c_r - c_{\rm F814W})\nonumber \\ 
              &   & - 7.5\log (1+z)\,,
\label{eq:ct}
\end{eqnarray}

\noindent
where the first term of the right member is the observed surface brightness listed in Table \ref{tab:lensphot}; the second one is the colour term between the redshifted Gunn r filter and the F814W passband, expressed in the AB magnitude system (Oke 1974\cite{oke74}) to measure absolute flux densities; the third term relates Vega to AB magnitudes; the last term corrects for the $(1+z)^3$ cosmological dimming for flux densities expressed in frequency units as in the AB system. The colour term has been derived by integrating the properly redshifted spectroscopic template of elliptical galaxies (Kinney et al. 1996\cite{kinney96}, Mannucci et al. 2001\cite{mannucci01}). $c_r$ and $c_{\rm F814W}$ have been obtained from Frei and Gunn (1994\cite{frei94}) and the ACS/WFC zeropoints\footnote{STScI Web site: http://www.stsci.edu/hst/acs/analysis/zeropoints} respectively. Similar relations are derived to compute the $B_z$ and $J_z$ rest frame magnitudes.

We found the following colour indices: $(B-r)_z = 1.27 \pm 0.13$ and $(B-J)_z = 3.17 \pm 0.14$. These values are similar to the colours of local elliptical galaxies ($(B-r)=1.17$, Frei \& Gunn 1994\cite{frei94}; $(B-r)=1.2$, van de Ven et al. 2003\cite{vandeven03} and $(B-J)=3.34\pm0.12$, Mannucci et al .2001\cite{mannucci01}; $(B-J)=3.25$, Fioc and Volmerange 1997\cite{fioc97}), with no sign of significant reddening.  Adding the fact that the surface brightness is estimated in the r band, we conclude that the value  $\mu_{e,r}(z=0)=21.11\pm 0.1$ should not be overestimated by dust extinction.

\begin{figure*}
\begin{center}
\hbox to \columnwidth{
\includegraphics[width=8cm]{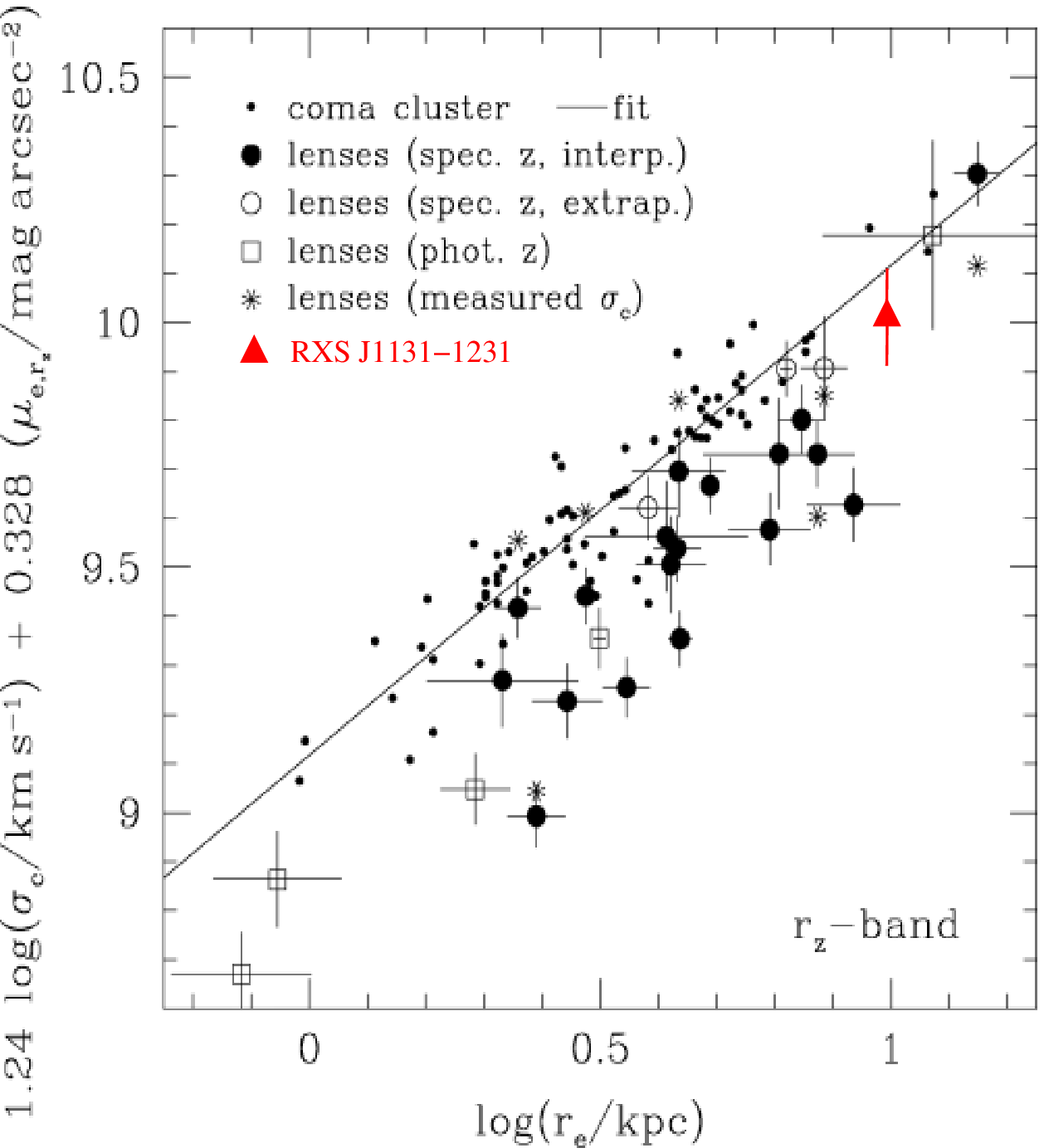}
\hfill
{\includegraphics[width=9cm]{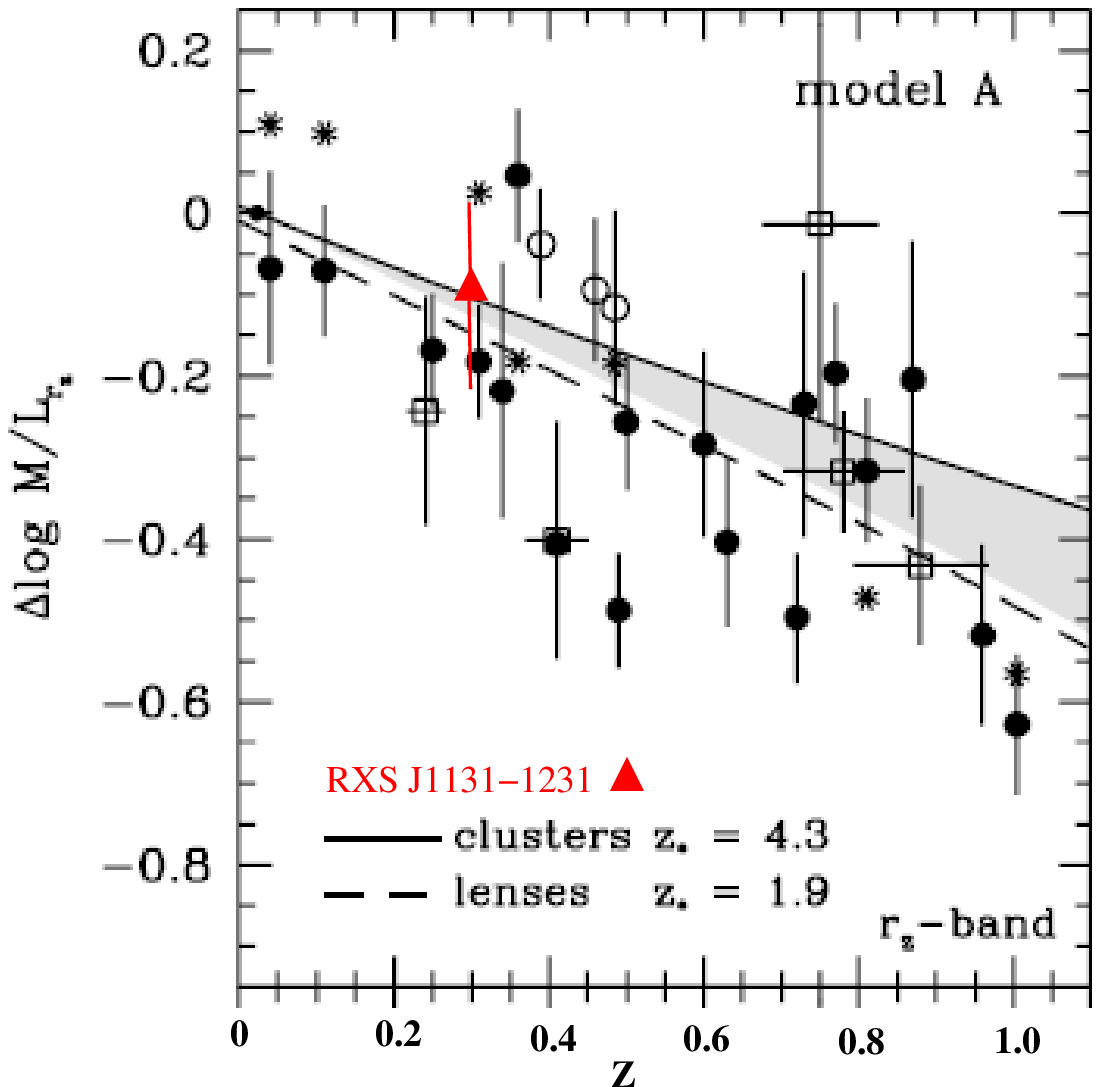}}
}
\caption{Left: Local fundamental plane in the rest-frame Gunn r-band, defined by the linear fit (J\o rgensen et al. 1995a\cite{jorg95a},b\cite{jorg95b}; thick line) of the Coma cluster galaxies at $z=0.023$ (small dots). The lens RXS J1131-1231 (triangle) is added to the lenses analysed by van de Ven et al. (2003\cite{vandeven03}).  Right: $M/L$ evolution in the Gunn r-band predicted for a single burst of stellar population occuring at $z_*=4.3$ (mean cluster formation; thick line) down to $z_*=2.0$ (maximum progenitor bias correction;shaded region) or at $z_*=1.9$ (best fit model for the lens galaxies analysed by van de Ven et al. 2003; dashed line). RXS J1131-1231 is added as a triangle.  The error bars on the abscisses are included in the symbol size; the errors on the ordinates reflect the dependence on the assumed value of $\sigma_c$ (see Tab. \ref{tab:ml}). Figures reproduced from van de Ven et al. (2003).}
\label{fig:fp}
\end{center}
\end{figure*}

\vspace{0.5cm}
\noindent
{\large {\it ii) $\sigma_c$: the Faber-Jackson relation and the lens environment}}
\vspace{0.25cm}

Lacking a high resolution, high S/N spectrum of the lens, the central velocity dispersion $\sigma_c$, needed in Eq. \ref{eq:fp}, must be derived either from the Faber-Jackson (FJ) relation (Faber \& Jackson 1976\cite{faber76}) or from the velocity dispersion of the best fitting isothermal ellipsoid, $\sigma_{\rm SIE}$. Ideally, both estimates should be equal.

Adopting the expression of the FJ relation given by Forbes \& Ponman (1999\cite{forbes99}), and an intrinsic scattering of $\Delta \log \sigma_c = 0.05$, we found $\sigma_{c,FJ} = 251\pm29$ km/s.

 On the other hand, $\sigma_{\rm SIE}$  can be derived from the fitted value of the Einstein radius $\theta_E$ according to the following relation:

\begin{equation}
\sigma_{\rm SIE} = 186.4 \sqrt{\frac{\theta_E('')D_{\rm OS}}{D_{\rm LS}}} \,\,{\rm km/s}\,,
\label{eq:re}
\end{equation}

\noindent
where $D_{\rm OS}$ (resp. $D_{\rm LS}$) is the angular distance from the observer (resp. the lens) to the source. From Sect. \ref{sec:fit5}, a robust estimate of $\theta_E = 1.84\pm0.01 ``$ is obtained and a value of $\sigma_{\rm SIE} = 355\pm 1$ km/s is derived.

The normalization factor $f_{\rm SIE} =  \sigma_c/\sigma_{\rm SIE}$ can be estimated theoretically under several hypotheses (e.g. assuming that the true profile {\it is really} isothermal). Kochanek (1994\cite{koch94}, 1996\cite{koch96}) and van de Ven et al. (2003\cite{vandeven03}) found that the expected value of the normalisation factor is compatible with 1, with a typical uncertainty of 10\%. However, recent spectroscopic observations of gravitational lenses show that $f$ might  rather be smaller than 1. Indeed, with a sample of five lenses, Treu and Koopmans (2004\cite{treu04}) found $f_{\rm SIE}=0.87\pm0.08$ (RMS) and in the case of B2045+265, Hamana et al. (2005\cite{hamana05}) even found $f=0.53$. Adopting the value $f=0.87\pm0.08$, we derived  $\sigma_{c,GL} = 309\pm 28$ km/s. Although  the values of $\sigma_{c,GL}$ and $\sigma_{\rm c,FJ}$ are quite different, the 1-$\sigma$ error bars nearly overlap.

However,  $\sigma_{\rm SIE}$ might also have been overestimated due to the presence of external convergence originating in the lens environment. By means of simulations, Keeton and Zabludof (2004\cite{keeton04}) have shown that typical  lens environment can produce a fractional overestimate of  $\sigma_{\rm SIE}$ by about 6\%. Correcting also for this effect would lead to the value $\sigma_{c,GL} = 291\pm 28$ km/s in better agreement with the FJ relation. A fractional increase of $\sigma_{\rm SIE}$ by 23\% due to the environment would be necessary to reproduce the value of $\sigma_{\rm c,FJ}$. This would require a dense matter sheet with $\kappa \sim 0.34$, which could be provided by associated and/or foreground galaxy clusters. Indeed, Williams et al. (2006\cite{williams06}) have identified two Red Sequences in the field, corresponding to galaxy groups or clusters in the vicinity of J1131, one at the lens redshift (offset $\sim 37''$) and one at $z\sim 0.1$ (offset $\sim 137''$, also seen in X-rays, Chandra OBSID 4814, PI: Claeskens, public data).

\vspace{0.5cm}
\noindent
{\large {\it iii) The lens evolution}}
\vspace{0.25cm}

 We have adopted the structural parameters of the best de Vaucouleurs model, i.e. $R_e = 9.3 \pm 0.2$ kpc (see Table \ref{tab:lensphot}), the derived value of $\mu_{e,r}(z=0)=21.11\pm 0.1$ (see above) and several estimates of $\sigma_c$ in the range 251 - 355 km/s (depending on whether the FJ relation or the GL model is used; see above). We then made use of Eqs. (\ref{eq:fp}) and (\ref{eq:ml}), to locate the lens with respect to the local FP and to estimate its evolution in the Gunn-r band (see left and right panel of  Fig. \ref{fig:fp}, respectively). 
\begin{table}[t]
\begin{center}
\caption{Observed evolution of the lens, as a function of the underlying assumption on $\sigma_c$: FJ or GL model with possible correction for central velocity dispersion  (f$_{\rm SIE}$) and typical environment (Env).}
\label{tab:ml}
\begin{tabular}{ccc}
\hline
 Hypothesis & $\sigma_c$ (km/s) & $\Delta \log (M/L)_r$ \\
\hline
GL & $355\pm35$ & $0.01\pm 0.06$ \\
GL+f$_{\rm SIE}$ & $309\pm28$ & $-0.1\pm0.08$\\
GL+f$_{\rm SIE}$+Env & $291\pm26$& $-0.14\pm 0.08$\\
FJ & $251\pm 29$& $-0.24\pm 0.08$\\
\hline
\end{tabular}
\end{center}
\end{table}

 More quantitatively, the evolution in the Gunn-r band of the M/L ratio is reported in Tab. \ref{tab:ml}. The evolution obtained with the normalized and environment corrected lens model best agrees with  the value $\Delta \log (M/L)_r = -0.14 \pm 0.03$, derived from the evolution rate $\Delta \log (M/L)_r/\Delta z = -0.47 \pm 0.11$ of lenses found by van de Ven et al. (2003\cite{vandeven03}).

However, except under the hypothesis that the correct value of $\sigma_c$ is derived from the FJ relation, for which a stronger evolution is found, no estimate can rule out the absence of evolution to better than $\sim 1.5 \sigma$ level.

In any case, the apparent evolution of the lens with respect to local elliptical galaxies is weak since:

\begin{enumerate}
\item
the rest frame colours of the lens,  $(B-r)_z$ and $(B-J)_z$ are similar to the ones of local ellipticals (see above);

\item
the total $M/L_{{\rm B}}$ within the effective radius is not significantly smaller than in local galaxies. Indeed, first assuming an isothermal mass distribution with $\sigma_{\rm SIE}=335$km/s (after a 6\% reduction due to the lens environment, Keeton \& Zabludof 2004\cite{keeton04}, see above), the total mass inside the effective radius is found to be $M = 3.9 \pm 0.7 \, 10^{11}\, h^{-1}_{50}$ M$_\odot$. Second the light: integrating the S\'ersic model within $R_e$ in the reference frame B$_z$ band and adopting $M_{\rm B,\odot}=5.47$ (Allen 1999\cite{allen99}), we find $L_{{\rm B}} = 4.96\pm0.5 \,10^{10}\, h_{50}^{-2}$ L$_\odot$ and $M/L_{{\rm B}} = 7.8 \pm 1 \,h_{50}$ (M/L$_{\rm B}$)$_\odot$. This is quite close to the {\it upper} limit observed within the effective radius of local elliptical galaxies ($3.5 < M/L_{\rm B}h_{50} < 8.1$, Gerhard et al. 2001\cite{gerhard01}).

\end{enumerate}

In summary, although, within the uncertainties on $\sigma_c$, the FP analysis shows a possible weak evolution of the lensing galaxy, the latter looks quite similar to local elliptical galaxies, in terms of rest frame colours and M/L ratio. This is not unexpected, firstly because a significant scatter is observed between individual objects (e.g. van de Ven et al. (2003\cite{vandeven03}) in their study of a sample of lensing galaxies). Secondly, massive  galaxies like this lens are observed to evolve more slowly with redshift (e.g. van der Wel et al. 2005\cite{vanderwel05}). On the other hand, the rich lens environment probably helps to reconcile the value of $\sigma_c$ derived from $\sigma_{\rm SIE}$ and the one expected from the FJ relation.

 A spectroscopic measurement of the lens velocity dispersion will of course provide a more definitive answer to this issue.

\begin{figure*}
\begin{center}
\includegraphics[width=15cm]{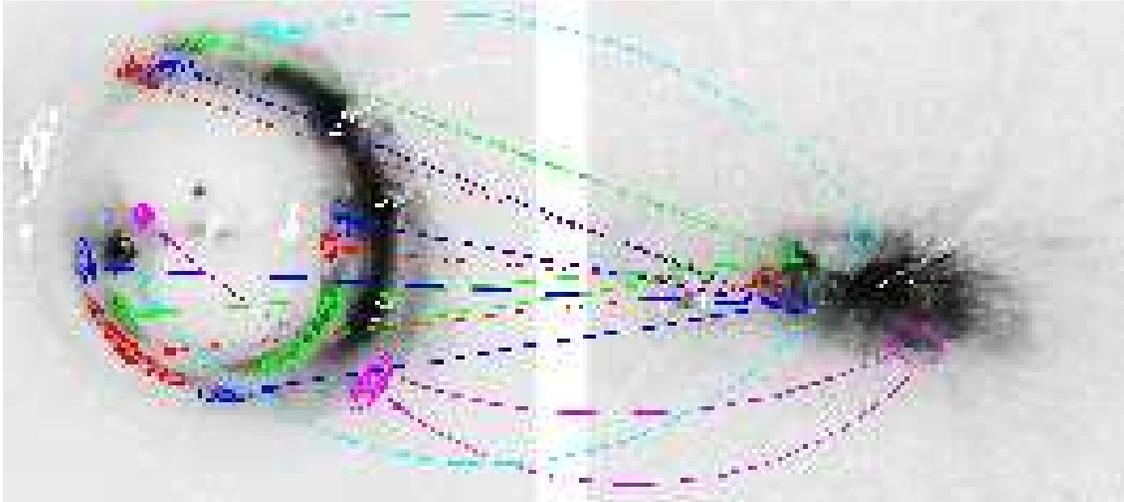}
\caption{Associations of the blue arcs and arclets in the image plane with the individual blue emitting regions in the source plane, based upon the fiducial lens model (illustrated in F814W).}
\label{fig:ident}
\end{center}
\end{figure*}

\subsection{The host galaxy}
\label{sec:host}

Visual inspection of the multicolour reconstucted host galaxy (see Fig. \ref{fig:source_rgb}) immediately reveals several striking details:

\begin{enumerate}
\item
Blue spots are easily located. These are associated with the blue arcs and arclets seen in the image plane (see Fig. \ref{fig:rgb}). The associations are indicated in Fig. \ref{fig:ident}. We identify one quadruplet of arclets, two triplets (with one pair of merging images in each of them) and three pairs (for which the identifications are made from a circular sub-source extracted from the real source). 

\item
The blue spots are spread along a spiral arm. Keeping in mind that the F555W filter roughly corresponds to the U band at $z=0.66$, we may infer that they are very likely associated with intense stellar formation in the host.

\item
On the left side of the host, a faint, blueish companion is clearly seen and is likely in interaction with the host, which indeed shows disturbances rather than a smooth profile.

\item
A strong radial colour gradient is seen across the host galaxy. It is not due to artifacts related to the PSF subtractions, since it is already visible in the Einstein ring of Fig. \ref{fig:rgb}. The relatively bluer (coded as yellow) central emission could be due to the presence of many young stars associated with a starburst in a nuclear ring (Knapen 2005\cite{knapen05}). This could be tested with 2-D spectroscopy.

\end{enumerate}

The following results can be deduced. First of all, statement number 1 underlines the coherence of our fiducial GL model since a correct association between the various arcs and arclets is possible. 

Secondly, the simultaneous observation of the AGN phenomenon, of stellar formation and of galaxy interaction illustrates the current AGN - black hole paradigm, according to which the growth of a black hole through the AGN phase occurs when  galaxy interactions provide gas inflow towards the center (e.g. Hernquist 1989\cite{hern89}, Di Matteo et al. 2005\cite{dimatteo05}). Since the interactions induce gravitational perturbations, simultaneous starbursts are also expected.  However, note that even if in the past, companions were found more frequently around active galaxies (e.g. Dahari 1984\cite{dahari84}), it does not seem to be systematic, nor even statistical (e.g. S\'anchez et al. 2004\cite{sanchez04}, Knapen 2005\cite{knapen05} and references therein).

Finally and more quantitatively, the source of the lens system J1131 must be a {\it luminous Seyfert 1 spiral} galaxy. This result relies on the following line of arguments:

\begin{enumerate}
\item
The host is a spiral, disk dominated galaxy, as derived from Fig. \ref{fig:source_rgb} and the reasonable fit by a S\'ersic profile (see Table \ref{tab:host} and Sect. \ref{sec:fit6}). This is corroborated by the rest frame index (U-V)$_z$ = $0.78\pm0.65$, derived from a transformation of the F555W and F814W measurements similar to Eq. \ref{eq:ct}. Such a colour index is typical of local spirals and excludes elliptical galaxies at $1 \sigma$ (Fioc \& Rocca-Volmerange 1997\cite{fioc97}).
\item
The rest frame absolute magnitude of the AGN is $M_{\rm V,z}({\rm AGN}) > -22.9 + 5\log h_{50}$. This is a robust upper limit estimated by adopting the flux of the brightest point-like image (i.e. image B) and assuming a total magnification of 10 (i.e. a conservative value given the macro lens model of the cusp configuration). Adopting the Einstein - de Sitter cosmological model would even lower the upper limit by about 0.5 mag. \item
One ``coherence argument'' is that even if only $\simeq$ 20\% of Seyferts are late-type galaxies (e.g. S\'anchez et al. 2004\cite{sanchez04}), disk dominated galaxies are not observed to host AGN more luminous than $M_{\rm v} \simeq -23$ (Dunlop et al. 2003\cite{dunlop03}).
\item
The rest frame absolute magnitude of the host is found to be $M_{V,z}$(host)$ = -23.5 \pm 0.2 + 5\log h_{50}$. This is about 1 magnitude more luminous than the most luminous Seyfert galaxy in Smith et al. (1986\cite{smith86}). However, a significant part (i.e. $\sim 0.5$ mag)  of this luminosity excess could be due to nebular emission associated with intense star formation activity. Indeed, [O{\sc iii}] $\lambda$5007, [O{\sc iii}] $\lambda$4959 and H$_\beta$ fall in the F814W filter and could also be responsible for the colour gradient mentioned above. After correction for this effect, this host galaxy would remain among the most luminous Seyfert galaxies.

\end{enumerate}

\section{Conclusions}
\label{sec:conclusion}

This paper was devoted to a thorough analysis of high angular resolution HST optical and NIR direct imagery of the complex gravitational lens system \object{RXS~J1131-1231}. 

We have provided precise astrometry and photometry in the F555W, F814W and F160W passbands for each component of the system, namely the 4 point-like lensed images of the central AGN, the lensed host galaxy, the lensing galaxy and the so-called X-object projected close to the lens. We have also reconstructed the image of the host in the source plane in a {\it non-parametric} way, leading to the view of an active galaxy at $z=0.658$ with unprecedented details.

The summary of our results is the following:

\begin{enumerate}
\item
A fiducial lens model consisting of a Singular Isothermal Ellipsoid with external shear perfectly reproduces the relative astrometry of the point-like lensed images. The shear direction is found to be compatible (within large error bars) with the position of a galaxy group identified at the lens redshift by Williams et al. (2006\cite{williams06}). However, a disagreement of $\sim 0.04''$ (i.e. $\sim 14 \sigma$) is found between the observed and predicted astrometric positions of the lens. This issue could be solved by adding to the fiducial model an $m=4$ multipole leading to boxy/disky surface mass density isophotes. The latter model simultaneously reproduces the observed 'anomalous' flux ratio $I_{\rm B}/I_{\rm C}$, confirming that lensing by substructures in the halo may not be required (see also Paper I). However, the physical nature of the fitted octupole is difficult to interpret and, as usual, no claim is made about the unicity of the proposed GL model.
\item
Faint but significant residuals are found when modelling the lens photometric profile in the F814W band with a S\'ersic profile, even when allowing for boxy isophotes. A slight offset with respect to the positions in the other passbands is also observed,  although no trace of dust is seen in the $(B-K)_z$ index. Would this be a trace of past interaction, maybe due to the  X-object? Are both issues 1- and 2- related? High angular resolution 2D spectroscopy is probably required to answer those questions. It is however striking that this kind of residuals in the ACS F814W band have also been reported for other lenses (Bolton et al. 2006\cite{bolton06}). 
\item
 The different values of $\sigma_c$ derived from lensing and from the FJ relation indicate that the contribution of the lens environment could be important. The latter looks indeed quite rich (Williams et al. 2006\cite{williams06}, Chandra observations, PI: Claeskens, public data).
\item
 Although, within the uncertainties on $\sigma_c$, the FP analysis shows a possible weak evolution of the lensing galaxy, the latter looks quite similar to local elliptical galaxies, in terms of rest frame colours and M/L ratio. Presence of dust is not detected. 
\item
The X-object is identified as possibly being a satellite galaxy of the lens, but with very little effect on the lens model. In any case, it is not the fifth lensed image of the AGN  nor a lensed substructure of the host.
\item
The reconstructed host is identified as a {\it luminous Seyfert 1 spiral} galaxy showing patchy emission in the UV rest frame, probably corresponding to intense star formation. A blueish closeby companion is also detected. The global shape and magnitudes of the host do not strongly depend on the addition of the octupole in the lens potential.
\item
This patchy UV emission is lensed into multiple thin arcs and arclets in the F555W band, whose positions are rather well reproduced by our lens models.
\item
In the peculiar case of the present cuspy lens system, extended structures did not turn out to be very helpful to constrain more general lens models. This underlines the need for robust astrometric constraints and this indicates that care must be taken when only relying on extended images to constrain a lens model.
\end{enumerate}

\begin{acknowledgements}

 We thank the anonymous referee for his constructive comments and interesting suggestions.  DS wants to thank P. Saha for useful discussion on PixeLens and lens modeling. We also thank R. Hook and the HST support team for their help about the ACS distortion correction. Our research was supported in part by {\sc Prodex}-HST (JFC, DS, JS), by contract IUAP P5/36 ``P\^ole d'Attraction Interuniversitaire'' (PR) and by the Swiss National Science Fundation (DS).

\end{acknowledgements}

\end{document}